\documentclass[aps,prm,amsmath,amssymb,reprint,superscriptaddress,]{revtex4-2}
\usepackage{charter,graphicx,verbatim,threeparttable,float,amssymb,gensymb }
\usepackage{upgreek}
\begin{document}

\preprint{APS/123-QED}

\title{YbV$_3$Sb$_4$ and EuV$_3$Sb$_4$, vanadium-based kagome metals with Yb$^{2+}$ and Eu$^{2+}$ zig-zag chains}

\author{Brenden R. Ortiz} 
\email{ortiz.brendenr@gmail.com}
 \affiliation{Materials Department, University of California Santa Barbara, Santa Barbara, CA 93106, USA}

 \author{Ganesh Pokharel}
 \affiliation{Materials Department, University of California Santa Barbara, Santa Barbara, CA 93106, USA}

\author{Malia Gundayao}
 \affiliation{Materials Department, University of California Santa Barbara, Santa Barbara, CA 93106, USA}%

\author{Hong Li}
 \affiliation{Department of Physics, Boston College, Chestnut Hill, MA 02467, USA}%
 
\author{Farnaz Kaboudvand}
 \affiliation{Materials Department, University of California Santa Barbara, Santa Barbara, CA 93106, USA}%

 \author{Linus Kautzsch}
 \affiliation{Materials Department, University of California Santa Barbara, Santa Barbara, CA 93106, USA}%

 \author{Suchismita Sarker}
 \affiliation{CHESS, Cornell University, Ithaca, NY, 14853, United States}%

\author{Jacob P. C. Ruff}
 \affiliation{CHESS, Cornell University, Ithaca, NY, 14853, United States}%

\author{Tom Hogan}
 \affiliation{Quantum Design, Inc., San Diego, CA 92121, USA}%
 
 \author{Steven J. Gomez Alvarado}
 \affiliation{Materials Department, University of California Santa Barbara, Santa Barbara, CA 93106, USA}%
 
 \author{Paul M. Sarte}
 \affiliation{Materials Department, University of California Santa Barbara, Santa Barbara, CA 93106, USA}%

 \author{Guang Wu}
 \affiliation{Department of Chemistry and Biochemistry, University of California Santa Barbara, Santa Barbara, CA 93106, USA}

 \author{Tara Braden}
 \affiliation{Physics Department, Colorado School of Mines, Golden, CO 80401, USA}
 
  \author{Ram Seshadri}
 \affiliation{Materials Department, University of California Santa Barbara, Santa Barbara, CA 93106, USA}
 
 \author{Eric S. Toberer}
 \affiliation{Physics Department, Colorado School of Mines, Golden, CO 80401, USA}

 \author{Ilija Zeljkovic}
 \affiliation{Department of Physics, Boston College, Chestnut Hill, MA 02467, USA}%
 
 \author{Stephen D. Wilson} \email{stephendwilson@ucsb.edu}
 \affiliation{Materials Department, University of California Santa Barbara, Santa Barbara, CA 93106, USA}%

\date{\today}

\begin{abstract}

 Here we present YbV$_3$Sb$_4$ and EuV$_3$Sb$_4$, two new compounds exhibiting slightly distorted vanadium-based kagome nets interleaved with zig-zag chains of divalent Yb$^{2+}$ and Eu$^{2+}$ ions. Single crystal growth methods are reported alongside magnetic, electronic, and heat capacity measurements. YbV$_3$Sb$_4$ is a nonmagnetic metal with no collective phase transitions observed between 60~mK and 300~K. Conversely, EuV$_3$Sb$_4$ is a magnetic kagome metal exhibiting easy-plane ferromagnetic-like order below $T_\text{C}=32$~K with hints of modulated spin texture under low field.  Our discovery of YbV$_3$Sb$_4$ and EuV$_3$Sb$_4$ demonstrate another direction for the discovery and development of vanadium-based kagome metals while incorporating the chemical and magnetic degrees of freedom offered by a rare-earth sublattice.
\end{abstract}

\maketitle

\section{Introduction}

Research into layered kagome metals has accelerated dramatically in the past few years, fueled in part by the discovery of the \textit{A}V$_3$Sb$_5$ kagome superconductors ~\cite{ortiz2019new,ortizCsV3Sb5,ortiz2020KV3Sb5,RbV3Sb5SC} and the continued exploration of the \textit{A}\textit{M}$_6$\textit{X}$_6$ phase space~\cite{PhysRevLett.127.266401,PhysRevB.103.014416,PhysRevB.104.235139,PhysRevMaterials104202,PhysRevB.103.014416,PhysRevB.106.115139,sciadv_abe2680,PhysRevLett.129.216402,Yin_2020,PhysRevMaterials.6.105001,PhysRevMaterials.6.083401,ZhangShao-ying_2001,PhysRevLett.126.246602}. Metals based on kagome networks derive much of their fundamental interest via their potential to realize an electronic structure replete with Dirac points, flat bands, and Van Hove singularities~\cite{park2021electronic,PhysRevB.87.115135,kiesel2013unconventional,meier2020flat}. Depending on the alignment of the Fermi level with the aforementioned features, a wide array of electronic instabilities ranging from bond density wave order ~\cite{PhysRevB.87.115135,PhysRevLett.97.147202}, charge fractionalization~\cite{PhysRevB.81.235115, PhysRevB.83.165118}, charge-density waves~\cite{PhysRevB.80.113102,yu2012chiral,kiesel2013unconventional}, and superconductivity~\cite{PhysRevB.87.115135,ko2009doped,kiesel2013unconventional} are possible. Developing new compounds built from kagome networks with variable band fillings and the ability to engineer additional interactions, such as magnetic order, remains an ongoing challenge.

The nonmagnetic kagome network of vanadium ions filled near the Van Hove points in the \textit{A}V$_3$Sb$_5$ (\textit{A}: K, Rb, Cs) class of kagome superconductors display a unique intertwining of charge density wave (CDW) order and a superconducting ground state~\cite{ortizCsV3Sb5,ortiz2020KV3Sb5,RbV3Sb5SC,ortiz2021fermi,zhao2021cascade,hu2022coexistence,kang2022microscopic,jiang2021unconventional}.  This renders them excellent platforms for exploring the electronic interactions on a kagome lattice. Interfacing the nonmagnetic vanadium kagome network with magnetic interstitial ions in a manner similar to \textit{Ln}V$_6$Sn$_6$ compounds remains a challenge. Addressing this challenge is motivated by the promise of engineering magnetic order within itinerant kagome metals to stabilize an interesting range of magnetic and electronic instabilities. These include, for instance, tunable Chern gaps~\cite{PhysRevLett.126.246602,Yin_2020,ZhangShao-ying_2001,PhysRevMaterials104202}, anomalous Hall effects~\cite{PhysRevB.103.014416,zhang2022magnetic,dhakal2021anisotropically}, and spin-charge coupled density waves~\cite{teng2022discovery}. 

One promising material class that deserves mention are metals of the form \textit{A}\textit{M}$_3$\textit{X}$_4$. These compounds exhibit slightly distorted \textit{M}-based kagome sublattices with zig-zag chains of \textit{A}-site ions. The potential for magnetism through choice of the \textit{A}-site provides a degree of chemical flexibility analogous to the \textit{A}\textit{M}$_6$\textit{X}$_6$ family. The \textit{A}\textit{M}$_3$\textit{X}$_4$ structures known to date are limited almost exclusively as \textit{A}Ti$_3$Bi$_4$ with (\textit{A}$^{3+}$: La$^{3+}$, Ce$^{3+}$, Sm$^{3+}$)~\cite{ovchinnikov2018synthesis,ovchinnikov2019bismuth}. The singular known exception is CaV$_3$Sb$_4$, where the Ti--Bi sublattice is swapped for V--Sb. This substitution is intriguing, and mirrors the reverse case of the \textit{A}V$_3$Sb$_5$ family, wherein the Ti--Bi variants RbTi$_3$Bi$_5$~\cite{werhahn2022kagome} and CsTi$_3$Bi$_5$~\cite{werhahn2022kagome,yang2022titanium} arose following the initial discovery of the V--Sb series~\cite{ortiz2019new}. The overall structure appears somewhat tolerant of several different \textit{A}-site valences, with both the trivalent rare-earths and divalent alkali-earth compounds (e.g. CaTi$_3$Bi$_4$, CaV$_3$Sb$_4$) known. However, to date, CaV$_3$Sb$_4$ remains the only known V--Sb \textit{A}\textit{M}$_3$\textit{X}$_4$~\cite{ovchinnikov2019bismuth}. Note that other non-stoichiometric antimonides like NdTi$_3$(Sb$_{0.9}$Sn$_{0.1}$)$_4$ are known~\cite{bie2007ternary}, and while these compounds do not exist at the purely antimonide limit, they suggest a degree of chemical tunability in \textit{A}\textit{M}$_3$\textit{X}$_4$ metals. Regardless, little is known about the physical properties of these materials.

In this work, we present the single crystal growth and characterization of two new \textit{A}\textit{M}$_3$\textit{X}$_4$ kagome metals: YbV$_3$Sb$_4$ and EuV$_3$Sb$_4$. Similar to the case of CaV$_3$Sb$_4$, both YbV$_3$Sb$_4$ and EuV$_3$Sb$_4$ are instances of divalent Yb$^{2+}$ and Eu$^{2+}$ $A$-site cations. Analogous to the \textit{A}V$_3$Sb$_5$ and \textit{Ln}V$_6$Sn$_6$ compounds, the vanadium sublattice appears nonmagnetic, leaving the magnetism to be dominated by the rare-earth element. As expected of Yb$^{2+}$, crystals of YbV$_3$Sb$_4$ are Pauli paramagnetic metals with no clear thermodynamic phase transitions from 60~mK to 300~K. In contrast, EuV$_3$Sb$_4$ exhibits a low-field ferromagnetic transition with a $T_\text{C}$ of approximately 32~K. The magnetism presents with an easy-plane anisotropy, and the susceptibility when $H\parallel c$ suggests a more complex magnetic ground state in the zero field limit (e.g. moment canting, helical states). Curie-Weiss and magnetic heat capacity analyses are consistent with magnetism originating from the full moment of $S=7/2$ Eu$^{2+}$ ions. Together our results continue to expand upon the known kagome metals, introducing new routes forward in the realization of complex electronic and magnetic ground states on kagome platforms.

\section{Experimental Methods}

\subsection{Single Crystal Synthesis}

YbV$_3$Sb$_4$ single crystals are grown through a self-flux method. Note that all handling of raw reagents, powders, and precursors was done within an argon-filled glove box with oxygen and water levels $<$1~ppm. A stoichiometric mixture is formed by combining Yb (rod, Alfa 99.9\%), V (powder, Sigma 99.9\%) and Sb (shot, Alfa 99.999\%) at a 1:3:4 ratio into tungsten carbide ball-mill vials. The vials are sealed under argon and milled for 3\,hr in a SPEX 8000D dual mixer/mill. Note that the milling proceeds in three 1~hr segments with intermediate (hand) grinding steps to dislodge agglomerates of Yb metal. As-received Yb rod was mechanically cleaned of any residual oxides before it was cleaved into $<$1~mm chunks. The as-received V powder was also cleaned through sonication in a mixture of EtOH and HCl. 

The resulting precursor powder can be annealed at 650\degree C to produce phase-pure polycrystalline powders of YbV$_3$Sb$_4$ (see ESI)\cite{ESI}. While we present some powder data in the ESI, note that all characterization within the main body of this manuscript was performed on single crystals. For single crystal methods, the as-milled precursor is loaded in 2\,mL high-density alumina (Coorstek) crucibles and sealed in carbon-coated fused silica ampules under $\sim$0.7\,atm of argon. The samples are heated to 1050\degree C at a rate of 200\degree C/hr before cooling to 800\degree C at a rate of 1-2\degree C/hr. Samples are then allowed to cool to room temperature before extracting single crystals mechanically from the solidified flux. Crystals are thin (100--500~$\upmu$m) hexagonal flakes with side lengths approximately 1--2~mm. The samples are a lustrous silver and can be exfoliated with some slight difficulty. 

It is worth noting that the compound does not melt congruently. The heating of the precursor phase above 1050\degree C causes the YbV$_3$Sb$_4$ powder to undergo peritectic decomposition. The resulting liquid phase subsequently acts as the flux upon cooling. Attempts to grow YbV$_3$Sb$_4$ through other fluxes (e.g. Sb, Bi, Sn, Pb) have not yet been successful. Further optimization of the flux conditions is still underway.

Single crystals of EuV$_3$Sb$_5$ are grown through a bismuth flux. A stoichiometric mixture of Eu (rod, Alfa 99.9\%), V (powder, Sigma 99.9\%), and Sb (shot, Alfa 99.999\%) is mixed with Bi (rod, Alfa 99.9999\%) at a 1:3:4:40 ratio. The elemental reagents are loaded in 2\,mL Canfield crucibles fitted with a catch crucible and a porous frit. The crucibles are sealed in carbon-coated fused silica ampules under $\sim$0.7\,atm of argon. The samples are heated to 1000\degree C at a rate of 200\degree C/hr before cooling to 400\degree C at a rate of 2\degree C/hr. The samples are centrifuged at 400\degree C to remove excess bismuth. Crystals are thin (10-50$\upmu$m) hexagonal flakes with side lengths approximately 100-250~$\upmu$m. The samples have a brilliant silver luster. 

\subsection{X-ray Diffraction}

The structure of YbV$_3$Sb$_4$ was first solved using powder diffraction data collected at the Advanced Photon Source (11-BM) with 0.412619\,\AA\, radiation (see ESI \cite{ESI}). Powders were sealed in Kapton capillaries after being diluted at a molar ratio of 1:4 with amorphous SiO$_2$ to reduce X-ray absorption. Structural solution was performed using charge flipping methods and the Topas V6 software package \cite{oszlanyi2004ab,oszlanyi2005ab,coelho2007charge,Coelho}. The composition of single crystals measured in this work were characterized using Energy Dispersive Spectroscopy (EDS) using a Hitachi TM4000 electron microscope equipped with an integrated Oxford Instruments EDS probe. Samples are stoichiometric within the error of unstandardized EDS.

Additional single-crystal measurements for both YbV$_3$Sb$_4$ and EuV$_3$Sb$_4$ were performed on a Bruker KAPPA APEX II diffractometer equipped with an APEX II CCD detector using a TRIUMPH monochromator with a Mo K$\alpha$ X-ray source ($\lambda$ = 0.71073~\AA). Furthermore, synchrotron X-ray diffraction experiments were carried out at the QM2 beam line at the Cornell High Energy Synchrotron Source (CHESS). The incident X-ray wavelength of $\lambda$ = 0.41328~\AA~ was selected using a double-bounce diamond monochromator. A cryostream of flowing helium was used for temperature control. The diffraction experiment was conducted in transmission geometry using a 6-megapixel photon-counting pixel-array detector with a silicon sensor layer. Data was collected in full 360\degree~ rotations with a step size of 0.1\degree. Scattering planes in reciprocal space were visualized using the NeXpy software package. The diffraction data was indexed and integrated using the APEX3 software package  including absorption and extinction corrections. Crystallographic structural solutions were determined using the SHELX software package \cite{sheldrick2008short}. 

\subsection{Computational Modeling}

As no prior electronic structure calculations are reported, we ventured to provide basic information about the electronic structure of the nonmagnetic \textit{A}\textit{M}$_3$\textit{X}$_4$ lattice. First-principles calculations based on density functional theory (DFT) within the Vienna ab initio Simulation Package (VASP) were performed \cite{kresse1996efficient,kresse1996efficiency}. The projector augmented wave (PAW) method \cite{blochl1994projector,kresse1999ultrasoft} was employed and relaxations of the ionic positions were conducted using an energy cutoff of 520 eV. Reciprocal space \textit{k}-point meshes were automatically generated at a density of 40~\AA$^{-1}$ along each reciprocal lattice vector. The band structure was calculated across a subset of the high symmetry points as defined by Setyawan and Curtarolo \cite{setyawan2010high}.

\subsection{Scanning Tunneling Microscopy}

Scanning tunneling microscopy (STM) measurements were performed using a customized Unisoku USM1300 microscope. Single-crystals of YbV$_3$Sb$_4$ were cleaved at 20~K and immediately inserted into the 4.5~K STM head. Spectroscopic measurements were made using a standard lock-in technique with 910~Hz frequency and bias excitation as detailed in figure captions. STM tips used were home-made chemically-etched tungsten tips, annealed in a UHV chamber to a bright orange color prior to the experiment. We apply the Lawler-Fujita drift-correction algorithm to all our data to align the atomic Bragg peaks onto single pixels.

\subsection{Bulk Characterization}

Magnetization measurements of both YbV$_3$Sb$_4$ and EuV$_3$Sb$_4$ single crystals were performed on a 7~T Quantum Design Magnetic Property Measurement System (MPMS3) SQUID magnetometer in vibrating-sample magnetometry (VSM) mode. Samples were mounted to quartz paddles using a small quantity of GE varnish. Two measurements were performed for both samples, orienting the \textit{c}-axis of the single-crystals perpendicular and parallel to the applied magnetic field. For consistency, the same crystal was used for both field orientations. 

Electronic resistivity measurements on YbV$_3$Sb$_4$ were performed on a Quantum Design 9~T Dynacool Physical Property Measurement System (PPMS). Single crystals were mounted to the sample stage using a small quantity of cigarette paper and GE varnish to ensure electrical isolation and thermal contact. Samples were then exfoliated and contacts established using silver paint (DuPont cp4929N-100) and gold wire (Alfa, 0.05~mm Premion 99.995\%). We used a dc current of 1~mA to measure the resistivity under zero-field conditions.

Heat capacity measurements on both YbV$_3$Sb$_4$ and EuV$_3$Sb$_4$ single crystals between 300~K and 1.8~K were performed on a Quantum Design 9~T Dynacool Physical Property Measurement System (PPMS) equipped with the heat capacity option. Measurements were repeated three times at each temperature and averaged together. Further measurements on YbV$_3$Sb$_4$ were performed on a Quantum Design 14~T Dynacool Physical Property Measurement System (PPMS) equipped with the dilution refrigerator (DR) option. Normalization of the DR heat capacity data was done through a scaling to higher temperature YbV$_3$Sb$_4$ data using the crossover from 2~K to 4~K.

\section{Results \& Discussion}

\subsection{Crystalline and Electronic Structure}

\begin{figure}
\includegraphics[width=\linewidth]{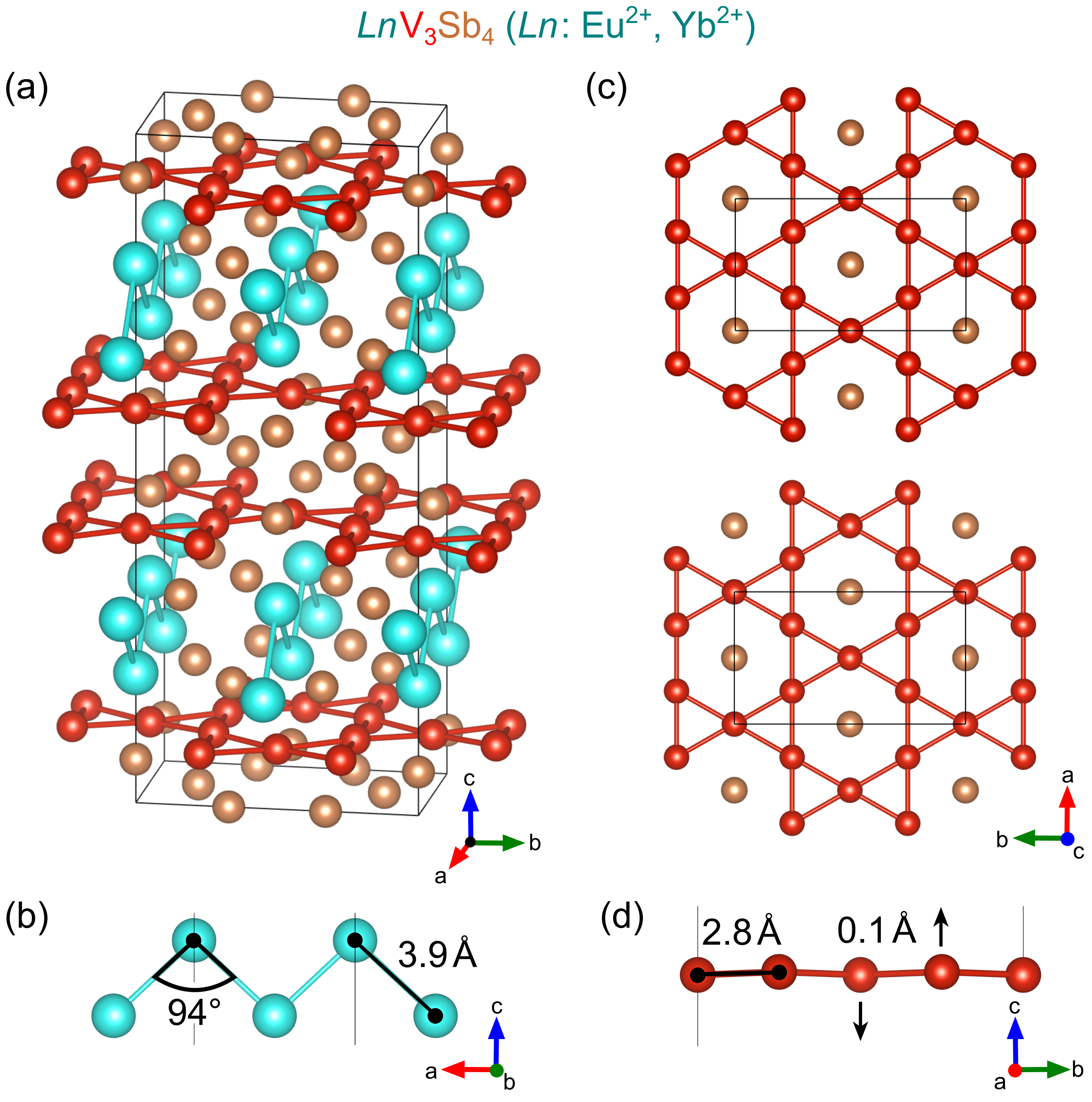}
\caption{YbV$_3$Sb$_4$ and EuV$_3$Sb$_4$ (a) are orthorhombic (\textit{Fmmm}) compounds that exhibit a zig-zag sublattice of \textit{Ln} ions (b) interwoven with staggered layers of V-based kagome networks (c). Consistent with the orthorhombic structure, the kagome networks are slightly distorted (d). The distortion is relatively minor, and all V-atoms are within 0.1~\AA~ of the idealized kagome lattice. Some interatomic distances of interest are highlighted on the graphic.}
\label{fig:1}
\end{figure}

\begin{figure*}
\includegraphics[width=1\textwidth]{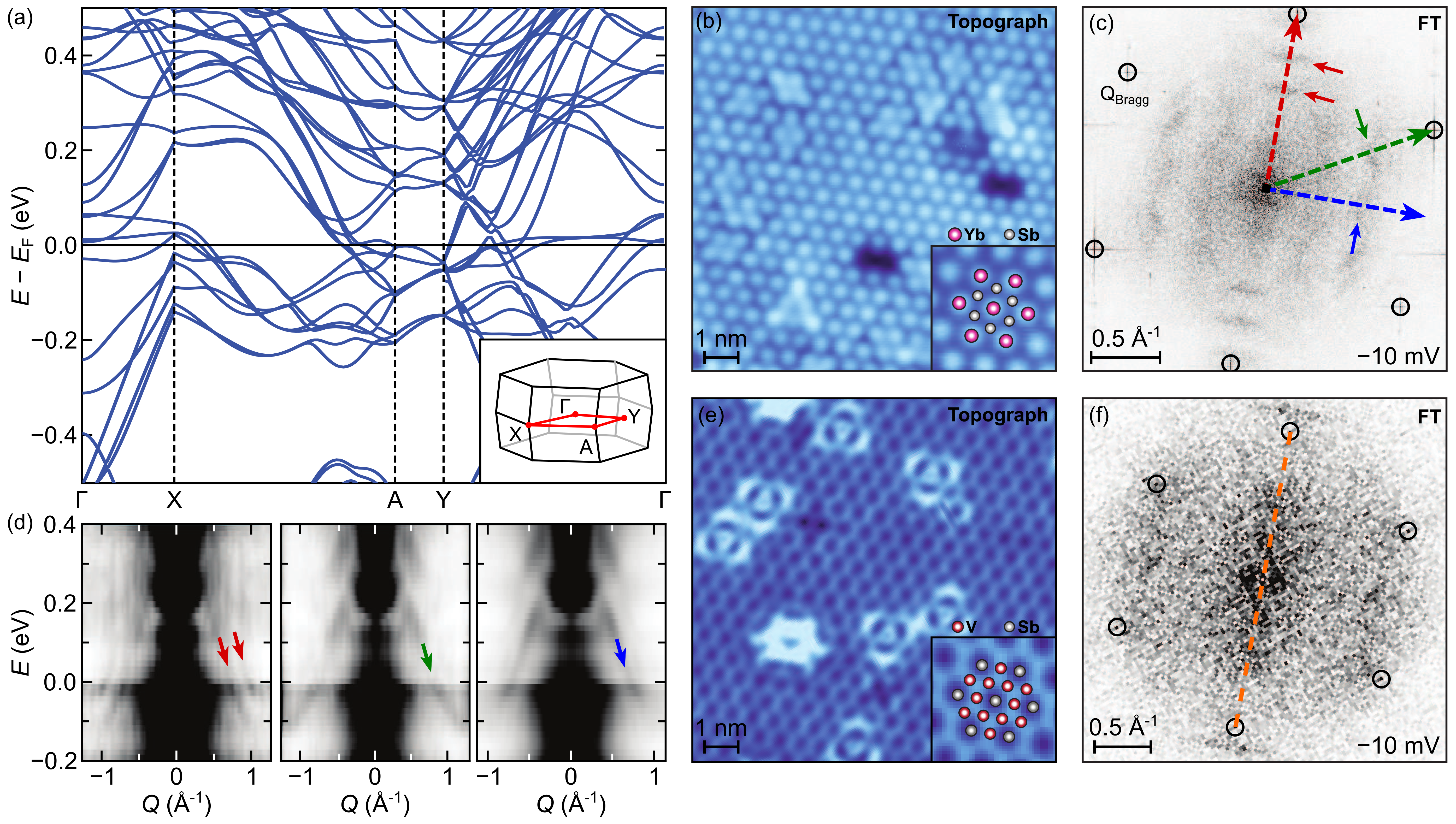}
\caption{(a) The electronic structure of \textit{Fmmm} YbV$_3$Sb$_4$ calculated over an abbreviated portion of the face-centered (type-1) orthorhombic high-symmetry points shows Dirac-like and flatband-like features consistent with the vanadium kagome network. Most cleavage surfaces exhibit Yb--Sb termination, resulting in the scanning tunneling microscopy (STM) topograph and quasiparticle interference (QPI) patterns shown in (b,c). Several energy- and momentum-dependent line cuts through the QPI patterns (red, green blue) are shown (d) with bands highlighted \textit{via} colored arrows. Though less common, crystals occasionally cleave along the V--Sb layers, resulting in the STM topograph shown in (e) and the corresponding QPI (f).}
\label{fig:2}
\end{figure*}

YbV$_3$Sb$_4$ and EuV$_3$Sb$_4$ are new members of a relatively small class of compounds \textit{A}\textit{M}$_3$\textit{X}$_4$. The majority of the members of this family consist of mildly distorted Ti--Bi networks with rare-earth cations \cite{ovchinnikov2019bismuth,motoyama2018magnetic,ovchinnikov2018synthesis}, though divalent calcium compounds (e.g. CaV$_3$Sb$_4$, CaTi$_3$Bi$_4$) were recently reported as well~\cite{ovchinnikov2019bismuth}. Unlike the known Ti--Bi rare-earth compounds, which can contain trivalent ions, both YbV$_3$Sb$_4$ and EuV$_3$Sb$_4$ form with rare-earth \textit{divalent} \textit{A}-site sublattices of Yb$^{2+}$ and Eu$^{2+}$. Figure \ref{fig:1}(a) illustrates the overall crystal structure of YbV$_3$Sb$_4$ and EuV$_3$Sb$_4$. We have chosen to omit the V--Sb and \textit{Ln}--Sb bonds to highlight the \textit{Ln} and V sublattices. Figure \ref{fig:1}(b) shows the \textit{Ln--Ln} distances, which can be visualized as zig-zag chains running in the \textit{a} direction. The \textit{Ln--Ln} distance along the chain ($\sim$3.9~\AA) is substantially closer than the nearest neighbor \textit{Ln-Ln} interchain distance ($\sim$5.6~\AA). 

Unlike the smaller prototype structures \textit{AM}$_3$\textit{X}$_5$ (\textit{P}6/\textit{mmm}) and \textit{AM}$_6$\textit{X}$_6$ (\textit{P}6/\textit{mmm}) compounds, the \textit{A}\textit{M}$_3$\textit{X}$_4$ (\textit{Fmmm}) compounds have four kagome layers in each unit cell. Figure \ref{fig:1}(c) highlights the different layers, which are offset from one another. For visual clarity we have left the nearest-neighbor Sb atoms displayed in Figure \ref{fig:1}(c), though they are not formally within the kagome plane. The case is analogous to the 
\textit{AM}$_6$\textit{X}$_6$ (HfFe$_6$Ge$_6$) prototype, where the \textit{X} atom is displaced slightly above/below the kagome sheet. In fact, the analogy with the \textit{AM}$_6$\textit{X}$_6$ compounds goes a bit further, as the \textit{A}\textit{M}$_3$\textit{X}$_4$ structure actually contains elements of the \textit{AM}$_6$\textit{X}$_6$ motif. If we consider stacking along the \textit{c}-axis, the \textit{A}\textit{M}$_3$\textit{X}$_4$ structure consists of $X_4$--$M_3$--$AX_2$--[$AX_2$--$M_3$--$X_4$--$M_3$--$AX_2$]--$AX_2$--$AX_2$--$M_3$--$X_4$ layers. The bracketed segment of the stacking represents the same stacking as the HfFe$_6$Ge$_6$ prototype structure. It has been noted that the other layers of  \textit{A}\textit{M}$_3$\textit{X}$_4$ compounds contain stacking elements drawing from many prototypical kagome and quasi-2D compounds (e.g. CeCo$_3$B$_2$, Cs$_2$Pt$_3$S$_4$, Zr$_4$Al$_3$, and CrSi$_2$) \cite{ovchinnikov2019bismuth}.

\begin{figure*}
\includegraphics[width=1\textwidth]{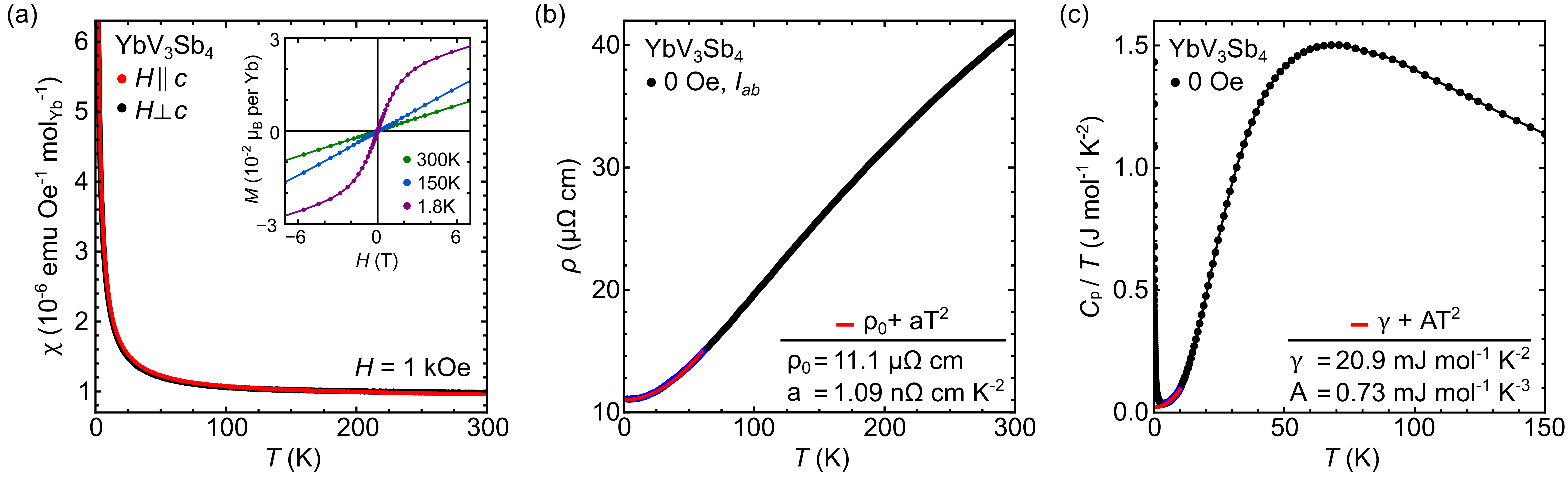}
\caption{Bulk electronic properties characterization on single crystals of YbV$_3$Sb$_4$. Temperature-dependent magnetization data are plotted in (a) with both orientations (\textit{c} parallel and perpendicular to \textit{H}) revealing largely temperature-independent Pauli paramagnetism with a weak Curie tail from impurity spins. Isothermal magnetization (a,inset) shows no clear saturation and a magnitude (10$^{-2}$ $\upmu_\text{B}$ per Yb) consistent with a small fraction of impurity spins. Zero-field resistivity measurements plotted in (b) confirm the metallic nature of YbV$_3$Sb$_4$ crystals. A simple quadratic fit to the low-temperature data is shown. Heat capacity results are plotted in (c). A nuclear Schottky anomaly is noted below 0.1~K. A Sommerfeld fit (red) is shown over a limited temperature range (blue) that avoids the Schottky anomaly.}
\label{fig:3}
\end{figure*}

Like other known  \textit{A}\textit{M}$_3$\textit{X}$_4$ compounds, the kagome layers in YbV$_3$Sb$_4$ and EuV$_3$Sb$_4$ are slightly distorted (see Figure \ref{fig:1}(d)). The planes are slightly buckled and the V--V distances are not identical. The distortions are small, however, and the vanadium atoms do not deviate more than 0.1~\AA~ from their idealized kagome positions. Corresponding CIF files have been included in the ESI \cite{ESI}. 

 Figure \ref{fig:2}(a) shows an abbreviated trace of the electronic structure near the Fermi level. An inset of the orthorhombic Brillouin zone showing a subset of the high-symmetry points is shown in the bottom right. Owing to the more complex structure, the band diagram near the Fermi level is substantially more involved than the \textit{AM}$_3$\textit{X}$_5$ and \textit{AM}$_6$\textit{X}$_6$ phases; however, a somewhat flat band feature from Y--$\Gamma$ seems to be close to the Fermi level. This is potentially accessible to future ARPES or scanning tunneling spectroscopy studies. 

STM measurements were used to screen YbV$_3$Sb$_4$ crystals for any short-range charge correlations not captured in the average structure. Crystals of YbV$_3$Sb$_4$ cleave in such a way that both the Yb--Sb and V--Sb terminations are available. Figure \ref{fig:2}(b) shows an STM topograph across a Yb--Sb cleavage plane. The lattice parameters observed by STM (\textit{a}=5.70~\AA~ and \textit{b}=9.87~\AA) agree well with those refined from diffraction (\textit{a}=5.62~\AA~ and \textit{b}=9.82~\AA). The associated Fourier transform and quasiparticle interference (QPI) patterns are also shown in Figure \ref{fig:2}(b-c), where the atomic Bragg peaks have been highlighted with black circles. The QPI is two-fold symmetric by inspection, consistent with the orthorhombic nature of YbV$_3$Sb$_4$. Several energy/momentum line cuts are highlighted on Figure \ref{fig:2}(c) in red, green, and blue. These cuts correspond to the energy- and momentum-dependent QPI linecuts shown in Figure \ref{fig:2}(d). Bands highlighted by colored arrows correspond directly to those highlighted in Figure \ref{fig:2}(c).

While substantially rarer, the V--Sb termination is also observed by STM. The corresponding topograph is shown in Figure \ref{fig:2}(e), and the QPI for the vanadium-terminated surface is shown in Figure \ref{fig:2}(f). The atomic Bragg peaks are outlined with black circles and the two-fold axis is highlighted in orange. A energy/momentum line cut through Figure \ref{fig:2}(f) is shown in the ESI \cite{ESI}. Due to the relatively complex electronic structure, more detailed calculations will be required to make direct comparisons between the bands extracted from the STM QPI linecuts and our DFT calculations.

\subsection{YbV$_3$Sb$_4$}

\begin{figure*}
\includegraphics[width=1\textwidth]{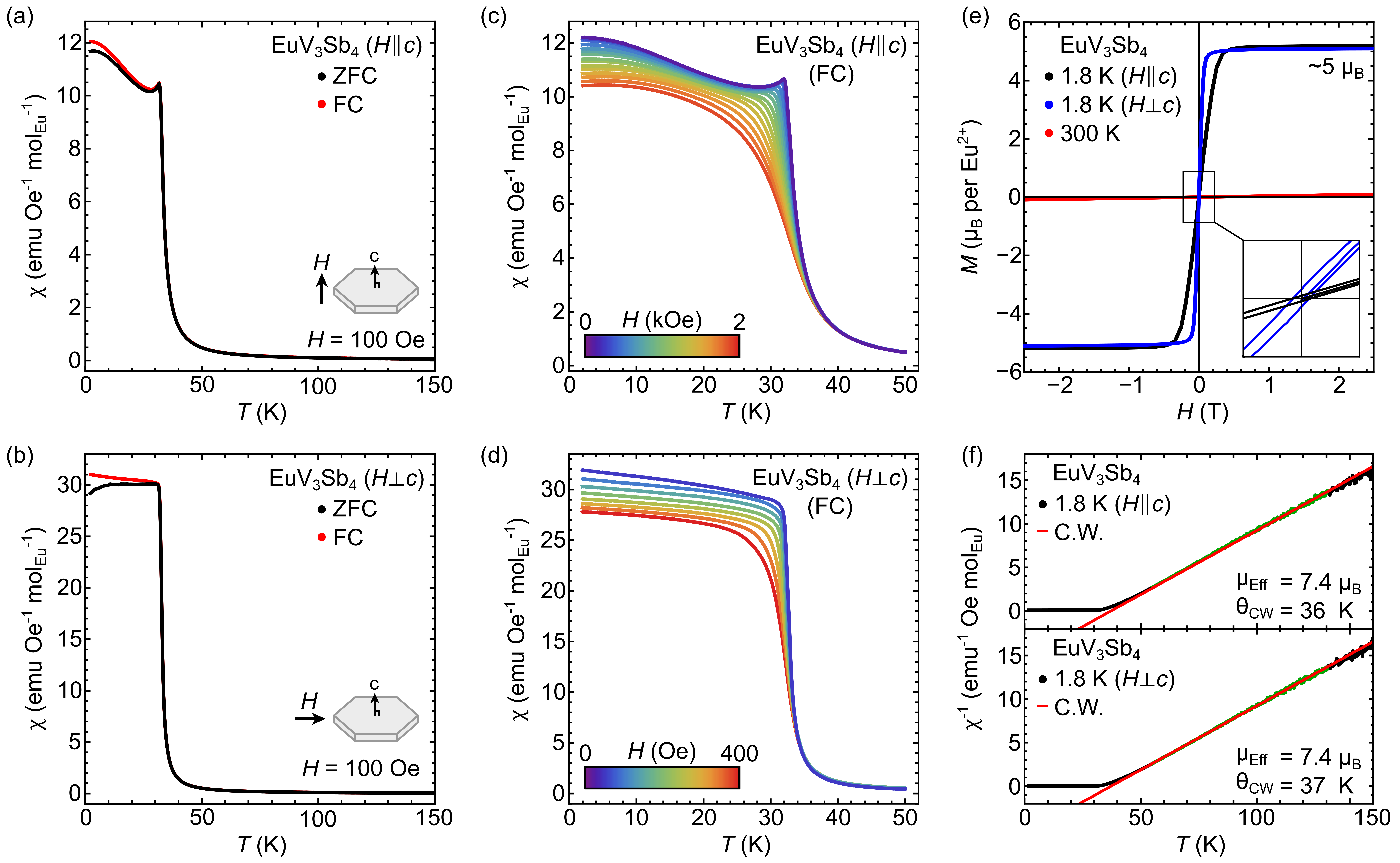}
\caption{Here we demonstrate a suite of magnetization data collected on single crystals of EuV$_3$Sb$_4$. Temperature-dependent susceptibility collected on a crystal oriented with \textit{H}$\parallel$\textit{c} (a) and \textit{H}$\perp$\textit{c} (b) both exhibit a sharp upturn in magnetization, consistent with ferromagnetic-like ordering near $T_\text{C}=32$~K. Note that an additional cusp and brief downturn in the susceptibility is noted when \textit{H}$\parallel$\textit{c}. Field-cooled measurements over a range of applied fields are plotted over the transition for  \textit{H}$\parallel$\textit{c} (c) and \textit{H}$\perp$\textit{c} (d). Isothermal ZFC magnetization data are plotted in (e) highlighting rapid moment saturation and a weak coercivity (inset). Curie-Weiss analysis of the low-field susceptibility for both field orientations are plotted in (f).} 
\label{fig:4}
\end{figure*}

We now turn to the characterization of the bulk electronic properties of YbV$_3$Sb$_4$ single crystals. Figure \ref{fig:3}(a) plots temperature-, field-, and orientation-dependent magnetization data. Evidenced by the extremely weak susceptibility (10$^{-6}$~emu~Oe$^{-1}$~mol$^{-1}$), YbV$_3$Sb$_4$ is a Pauli paramagnet with no signatures of bulk, local moments. No significant qualitative difference is noted when the crystal is mounted with the \textit{c}-axis parallel or perpendicular to the magnetic field. Similarly, isothermal magnetization on single crystals (Figure \ref{fig:3}(a,inset)) shows no saturation. Please note that the scale of the inset plot is on the order of 10$^{-2}$~$\upmu_\text{B}$ per Yb, consistent with the polarization of impurity spins. 

Temperature-dependent resistivity data showing metallic transport in YbV$_3$Sb$_4$ are plotted in Figure \ref{fig:3}(b). The residual resistivity is approximately 11~$\upmu\Omega$~cm, though the residual resistivity ratio (RRR) is relatively low ($\sim$4). As a result, no signatures of quantum oscillations in the magnetoresistance were observed with \textit{I}$_\text{ab}$ and \textit{H}$\parallel$\textit{c}. The zero-field, low temperature ($<50$~K) resistivity is well modeled via Fermi liquid behavior and a simple quadratic fit $\rho = \rho_0 + aT^2$ with $\rho_0 = 11.1~\upmu\Omega$~cm and $a=1.09$~n$\Omega$~cm~K$^{-2}$.

No clear phase transitions are indicated in the magnetization or resistivity results on YbV$_3$Sb$_4$ down to 2~K. Zero-field heat capacity were also collected from 300~K down to 60~mK. Figure \ref{fig:3}(c) plots the resulting $C_\text{p}/T$ data for a 1.1~mg single crystal.  A feature consistent with a nuclear Schottky anomaly emerges around 0.1~K, and another small feature is noted at 0.8~K (see ESI)~\cite{ESI}. The  magnitude of the 0.8~K feature suggests this is due to an impurity effect, such as freezing of the paramagnetic impurities resolved in magnetization measurements. To avoid contributions from the nuclear Schottky anomaly, a Sommerfeld model was fit to a limited temperature range (4~K to 10~K). The resulting least-squares fit yields the parameters $\gamma=20.9$ mJ~mol$^{-1}$~K$^{-1}$ and $A=0.73$ mJ~mol$^{-1}$~K$^{-3}$.

All experimental data characterizing YbV$_3$Sb$_4$ support the classification as a nonmagnetic kagome metal. The structure is consistent with a Yb$^{2+}$ rare-earth sublattice, and no bulk magnetic, electronic, or structural instabilities are noted from 60~mK to 300~K. These results resemble those seen in CaV$_3$Sb$_4$ and CaTi$_3$Bi$_4$~\cite{ovchinnikov2019bismuth}. As such, YbV$_3$Sb$_4$ provides an excellent comparison and nonmagnetic standard for EuV$_3$Sb$_4$. 

\subsection{EuV$_3$Sb$_4$}

Whereas Yb$^{2+}$ results in a nonmagnetic rare-earth sublattice, divalent Eu$^{2+}$ is isoelectronic to Gd$^{3+}$ ($S=7/2$) and should exhibit a magnetically ordered ground state. Figure \ref{fig:4} shows the temperature-, field-, and orientation-dependent magnetization data from a 10$~\upmu$g single crystal of EuV$_3$Sb$_4$. Looking first at the temperature-dependent susceptibility, Figure \ref{fig:4}(a,b) plots the low-field susceptibility of EuV$_3$Sb$_4$ under an applied field of 100~Oe oriented with \textit{H}$\parallel$\textit{c} and \textit{H}$\perp$\textit{c}, respectively.

Both orientations exhibit a dramatic increase in the magnetization near 36 K. With \textit{H}$\perp$\textit{c}, a rapid polarization is observed that quickly saturates and suggests a predominantly ferromagnetic transition. Upon changing the field orientation such that \textit{H}$\parallel$\textit{c}, two main differences arise: (1) the magnitude of the susceptibility is dramatically reduced, demonstrating an easy-plane anisotropy, and (2) a low-field, sharp cusp appears near $T_C$ before continuing toward saturation.  Similar low field cusps have been observed in other magnetic kagome metals such as GdV$_6$Sn$_6$, suggesting a modulated magnetic ground state \cite{PhysRevB.104.235139}.

\begin{figure*}
\includegraphics[width=1\textwidth]{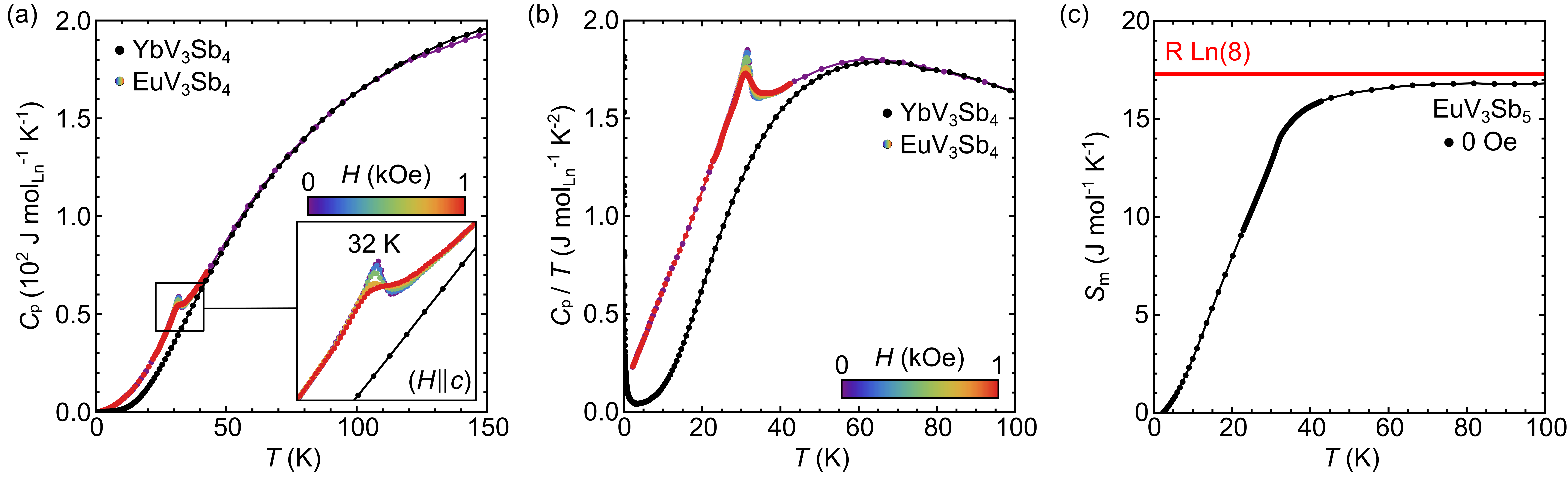}
\caption{Consistent with magnetization results, heat capacity measurements on single crystals of EuV$_3$Sb$_4$ clearly show a lambda-like anomaly at $T_\text{C}=32$~K. The transition is field-dependent (a,inset), broadening and weakly shifting towards lower temperatures. The transition is more obvious in $C_\text{p}/T$ (b), where direct subtraction of the scaled YbV$_3$Sb$_4$ nonmagnetic lattice reference yields the magnetic contribution to the total heat capacity. Integration of the magnetic heat capacity results in the magnetic entropy (c), yielding 97\% of $R\ln 8$, consistent with divalent Eu$^{2+}$.}
\label{fig:5}
\end{figure*}

Figure \ref{fig:4}(c,d) reports the field-dependent and temperature-dependent magnetization for both field orientations shown in Figure \ref{fig:4}(a,b). The low-field cusp in the magnetization for \textit{H}$\parallel$\textit{c} persists for fields up to approximately 500~Oe. The isothermal magnetization (Figure \ref{fig:4}(e)) demonstrates that the magnetic response of \textit{H}$\perp$\textit{c} saturates much quicker ($\sim$500~Oe) than \textit{H}$\parallel$\textit{c} ($\sim$2500~Oe), consistent with an easy-plane anisotropy. Figure \ref{fig:4}(e,inset) highlights the clear (albeit subtle) hysteresis in the 1.8 K isothermal magnetization, consistent with the FC/ZFC irreversibility plotted in Figure \ref{fig:4}(a,b). Curiously, at high fields, both curves saturate near 5$\upmu_\text{B}$, which is well below the expected value of 7~$\upmu_\text{B}$ for Eu$^{2+}$ assuming $S=7/2$ and $g=2$. Though Figure \ref{fig:4}(f) shows an abbreviated field range, no metamagnetic transitions were observed up to 7~T.

A Curie-Weiss analysis (Figure \ref{fig:4}(f)) of the low-field susceptibility in both field orientations produces nearly identical results. Both orientations exhibit $\uptheta_\text{CW}$ of approximately +36~K, consistent with both the heat capacity peak (32~K, discussed later) and the first derivative of the magnetization curve (33~K). The analysis supports predominantly ferromagnetic correlations and minimal frustration (recall that the \textit{Ln}-sublattice exhibits a zig-zag motif independent of the vanadium kagome lattice). The effective paramagnetic moment is 7.4~$\upmu_\text{B}$, which is close to that expected for Eu$^{2+}$ (7.9~$\upmu_\text{B}$).

It is worth noting that all results in Figure \ref{fig:4} were collected on the same crystal. We suspect that the local moment obtained via the Curie-Weiss analysis likely deviates from the full Eu$^{2+}$ moment due to uncertainty in determining the sample mass. An approximate 13\% error in mass would account for the deviation in the Curie-Weiss analysis, not outside expectations considering the small mass of the single crystal (10~$\upmu$g). However, the mass error still fails to account for the missing moment in the isothermal magnetization saturated state. Even compensated for the mass error, we approximate that $\approx$1 $\upmu_\text{B}$ remains unaccounted for, which may hint at unforeseen moment polarization on the vanadium sites or significant dynamic effects; however, future measurements on larger volume crystals will be required to completely rule out unaccounted massing errors. 

Figure \ref{fig:5} plots the heat capacity data for EuV$_3$Sb$_4$ utilizing YbV$_3$Sb$_4$ as a nonmagnetic phonon reference. A cluster of crystals weighing 0.5~mg were used for the heat capacity analysis. Figure \ref{fig:5}(a) shows heat capacity data overplotted for both compounds, including a small scaling factor that normalized the high-temperature heat capacity signals $>$100~K together. A clear anomaly can be seen in the heat capacity for EuV$_3$Sb$_4$ at $T_\text{C} = 32$~K, in good agreement with the magnetization results. We briefly highlight the field-dependence of the heat capacity anomaly in EuV$_3$Sb$_5$ when \textit{H}$\parallel$\textit{c}, demonstrating that the transition broadens and is slightly depressed in temperature with increasing fields. 

Figure \ref{fig:5}(b) shows the corresponding $C_\text{p}/T$ data for YbV$_3$Sb$_4$ and EuV$_3$Sb$_4$. Neglecting data near the nuclear Schottky anomaly in YbV$_3$Sb$_4$, direct subtraction of the nonmagnetic YbV$_3$Sb$_4$ lattice reference isolates the magnetic entropy from the EuV$_3$Sb$_4$ heat capacity data. Integration of $C_\text{p,mag}$ (see ESI) from 1.8~K to 100 K is plotted in Figure \ref{fig:5}(c), and the recovered entropy plateaus near 16.8~J~mol$^{-1}$~K$^{-1}$. This is 97\% of the entropy expected from Eu$^{2+}$ ions in the fully ordered state ($R\ln 8$). 

In aggregate, the data suggest that EuV$_3$Sb$_4$ adopts a noncollinear or modulated (e.g. helical or cycloidal) magnetic ground state in the zero field limit. This may account for the cusp in the magnetization immediately below $T_\text{C}$ observed in the hard-axis susceptibility data. However, larger samples and future scattering experiments (e.g., resonant x-ray or neutron scattering) will be required to fully explore this possibility.

\section{Conclusion}

We presented two new vanadium-based kagome materials, YbV$_3$Sb$_4$ and EuV$_3$Sb$_4$. These materials are members of a larger \textit{A}\textit{M}$_3$\textit{X}$_4$ family, which have generally consisted of Ti--Bi based kagome metals with rare-earth ions. Whereas the vanadium kagome sublattice appears nonmagnetic, the rare-earth sublattice creates zig-zag chains that form between kagome planes and introduce the potential for magnetic degrees of freedom.  YbV$_3$Sb$_4$ forms as a nonmagnetic kagome metal with no bulk phase transitions down to 60~mK, while EuV$_3$Sb$_4$ realizes a ferromagnetic-like ground state below $T_\text{C}=32$~K. The ordered state has an easy-plane anisotropy that shows hints of a canted or modulated order in the zero-field limit, motivating future exploration of its magnetic ground state. This work establishes the \textit{A}V$_3$Sb$_4$ class of kagome metals as new platforms for proximitizing \textit{A}-site tuned magnetic order with the topologically nontrivial features endemic to kagome band structures.

\section{Acknowledgments}

This work was supported by the US Department of Energy (DOE), Office of Basic Energy Sciences, Division of Materials Sciences and Engineering under Grant No. DE-SC0020305 (B.R.O., G.P., S.D.W.). B.R.O. and P.M.S. acknowledge financial support from the University of California, Santa Barbara, through the Elings Fellowship. A portion of this research, including the undergraduate internship program for M.G., F.K., and R.S. was supported by the National Science Foundation (NSF) through Enabling Quantum Leap: Convergent Accelerated Discovery Foundries for Quantum Materials Science, Engineering and Information (Q-AMASE-i): Quantum Foundry at UC Santa Barbara (DMR-1906325). I. Z. acknowledges support via NSF grant DMR 2216080. S.J.G.A. acknowledges financial support from the National Science Foundation Graduate Research Fellowship under Grant No. 1650114. A portion of this research, including support for E.S.T. and T.B., was supported by the National Science Foundation (NSF) under Grants No. DMR-1950924 and No. DMR-1555340, respectively. This work is based upon research conducted at the Center for High Energy X-ray Sciences (CHEXS), which is supported by the National Science Foundation under Award DMR-1829070. Use of the Advanced Photon Source at Argonne National Laboratory was supported by the U.S. Department of Energy, Office of Science, Office of Basic Energy Sciences, under Contract No. DE-AC02-06CH11357. The research made use of the shared experimental facilities of the NSF Materials Research Science and Engineering Center at UC Santa Barbara (Grant No. DMR-1720256). 

\bibliography{LnV3Sb4}

\providecommand{\noopsort}[1]{}\providecommand{\singleletter}[1]{#1}%
\begin{thebibliography}{51}%
\makeatletter
\providecommand \@ifxundefined [1]{%
 \@ifx{#1\undefined}
}%
\providecommand \@ifnum [1]{%
 \ifnum #1\expandafter \@firstoftwo
 \else \expandafter \@secondoftwo
 \fi
}%
\providecommand \@ifx [1]{%
 \ifx #1\expandafter \@firstoftwo
 \else \expandafter \@secondoftwo
 \fi
}%
\providecommand \natexlab [1]{#1}%
\providecommand \enquote  [1]{``#1''}%
\providecommand \bibnamefont  [1]{#1}%
\providecommand \bibfnamefont [1]{#1}%
\providecommand \citenamefont [1]{#1}%
\providecommand \href@noop [0]{\@secondoftwo}%
\providecommand \href [0]{\begingroup \@sanitize@url \@href}%
\providecommand \@href[1]{\@@startlink{#1}\@@href}%
\providecommand \@@href[1]{\endgroup#1\@@endlink}%
\providecommand \@sanitize@url [0]{\catcode `\\12\catcode `\$12\catcode
  `\&12\catcode `\#12\catcode `\^12\catcode `\_12\catcode `\%12\relax}%
\providecommand \@@startlink[1]{}%
\providecommand \@@endlink[0]{}%
\providecommand \url  [0]{\begingroup\@sanitize@url \@url }%
\providecommand \@url [1]{\endgroup\@href {#1}{\urlprefix }}%
\providecommand \urlprefix  [0]{URL }%
\providecommand \Eprint [0]{\href }%
\providecommand \doibase [0]{https://doi.org/}%
\providecommand \selectlanguage [0]{\@gobble}%
\providecommand \bibinfo  [0]{\@secondoftwo}%
\providecommand \bibfield  [0]{\@secondoftwo}%
\providecommand \translation [1]{[#1]}%
\providecommand \BibitemOpen [0]{}%
\providecommand \bibitemStop [0]{}%
\providecommand \bibitemNoStop [0]{.\EOS\space}%
\providecommand \EOS [0]{\spacefactor3000\relax}%
\providecommand \BibitemShut  [1]{\csname bibitem#1\endcsname}%
\let\auto@bib@innerbib\@empty
\bibitem [{\citenamefont {Ortiz}\ \emph {et~al.}(2019)\citenamefont {Ortiz},
  \citenamefont {Gomes}, \citenamefont {Morey}, \citenamefont {Winiarski},
  \citenamefont {Bordelon}, \citenamefont {Mangum}, \citenamefont {Oswald},
  \citenamefont {Rodriguez-Rivera}, \citenamefont {Neilson}, \citenamefont
  {Wilson} \emph {et~al.}}]{ortiz2019new}%
  \BibitemOpen
  \bibfield  {author} {\bibinfo {author} {\bibfnamefont {B.~R.}\ \bibnamefont
  {Ortiz}}, \bibinfo {author} {\bibfnamefont {L.~C.}\ \bibnamefont {Gomes}},
  \bibinfo {author} {\bibfnamefont {J.~R.}\ \bibnamefont {Morey}}, \bibinfo
  {author} {\bibfnamefont {M.}~\bibnamefont {Winiarski}}, \bibinfo {author}
  {\bibfnamefont {M.}~\bibnamefont {Bordelon}}, \bibinfo {author}
  {\bibfnamefont {J.~S.}\ \bibnamefont {Mangum}}, \bibinfo {author}
  {\bibfnamefont {I.~W.}\ \bibnamefont {Oswald}}, \bibinfo {author}
  {\bibfnamefont {J.~A.}\ \bibnamefont {Rodriguez-Rivera}}, \bibinfo {author}
  {\bibfnamefont {J.~R.}\ \bibnamefont {Neilson}}, \bibinfo {author}
  {\bibfnamefont {S.~D.}\ \bibnamefont {Wilson}}, \emph {et~al.},\ }\bibfield
  {title} {\bibinfo {title} {{New kagome prototype materials: discovery of
  KV$_3$Sb$_5$, RbV$_3$Sb$_5$, and CsV$_3$Sb$_5$}},\ }\href@noop {} {\bibfield
  {journal} {\bibinfo  {journal} {Phys. Rev. Materials}\ }\textbf {\bibinfo
  {volume} {3}},\ \bibinfo {pages} {094407} (\bibinfo {year}
  {2019})}\BibitemShut {NoStop}%
\bibitem [{\citenamefont {Ortiz}\ \emph
  {et~al.}(2020{\natexlab{a}})\citenamefont {Ortiz}, \citenamefont {Teicher},
  \citenamefont {Hu}, \citenamefont {Zuo}, \citenamefont {Sarte}, \citenamefont
  {Schueller}, \citenamefont {Abeykoon}, \citenamefont {Krogstad},
  \citenamefont {Rosenkranz}, \citenamefont {Osborn}, \citenamefont {Seshadri},
  \citenamefont {Balents}, \citenamefont {He},\ and\ \citenamefont
  {Wilson}}]{ortizCsV3Sb5}%
  \BibitemOpen
  \bibfield  {author} {\bibinfo {author} {\bibfnamefont {B.~R.}\ \bibnamefont
  {Ortiz}}, \bibinfo {author} {\bibfnamefont {S.~M.}\ \bibnamefont {Teicher}},
  \bibinfo {author} {\bibfnamefont {Y.}~\bibnamefont {Hu}}, \bibinfo {author}
  {\bibfnamefont {J.~L.}\ \bibnamefont {Zuo}}, \bibinfo {author} {\bibfnamefont
  {P.~M.}\ \bibnamefont {Sarte}}, \bibinfo {author} {\bibfnamefont {E.~C.}\
  \bibnamefont {Schueller}}, \bibinfo {author} {\bibfnamefont {A.~M.}\
  \bibnamefont {Abeykoon}}, \bibinfo {author} {\bibfnamefont {M.~J.}\
  \bibnamefont {Krogstad}}, \bibinfo {author} {\bibfnamefont {S.}~\bibnamefont
  {Rosenkranz}}, \bibinfo {author} {\bibfnamefont {R.}~\bibnamefont {Osborn}},
  \bibinfo {author} {\bibfnamefont {R.}~\bibnamefont {Seshadri}}, \bibinfo
  {author} {\bibfnamefont {L.}~\bibnamefont {Balents}}, \bibinfo {author}
  {\bibfnamefont {J.}~\bibnamefont {He}},\ and\ \bibinfo {author}
  {\bibfnamefont {S.~D.}\ \bibnamefont {Wilson}},\ }\bibfield  {title}
  {\bibinfo {title} {{CsV$_3$Sb$_5$: a $\mathbb{Z}_2$ topological kagome metal
  with a superconducting ground state}},\ }\href@noop {} {\bibfield  {journal}
  {\bibinfo  {journal} {Phys. Rev. Lett.}\ }\textbf {\bibinfo {volume} {125}},\
  \bibinfo {pages} {247002} (\bibinfo {year} {2020}{\natexlab{a}})}\BibitemShut
  {NoStop}%
\bibitem [{\citenamefont {Ortiz}\ \emph
  {et~al.}(2020{\natexlab{b}})\citenamefont {Ortiz}, \citenamefont {Kenney},
  \citenamefont {Sarte}, \citenamefont {Teicher}, \citenamefont {Seshadri},
  \citenamefont {Graf},\ and\ \citenamefont {Wilson}}]{ortiz2020KV3Sb5}%
  \BibitemOpen
  \bibfield  {author} {\bibinfo {author} {\bibfnamefont {B.~R.}\ \bibnamefont
  {Ortiz}}, \bibinfo {author} {\bibfnamefont {E.}~\bibnamefont {Kenney}},
  \bibinfo {author} {\bibfnamefont {P.~M.}\ \bibnamefont {Sarte}}, \bibinfo
  {author} {\bibfnamefont {S.~M.}\ \bibnamefont {Teicher}}, \bibinfo {author}
  {\bibfnamefont {R.}~\bibnamefont {Seshadri}}, \bibinfo {author}
  {\bibfnamefont {M.~J.}\ \bibnamefont {Graf}},\ and\ \bibinfo {author}
  {\bibfnamefont {S.~D.}\ \bibnamefont {Wilson}},\ }\bibfield  {title}
  {\bibinfo {title} {{Superconductivity in the $\mathbb{Z}_2$ kagome metal
  KV$_3$Sb$_5$}},\ }\href@noop {} {\bibfield  {journal} {\bibinfo  {journal}
  {Phys. Rev. Mater.}\ }\textbf {\bibinfo {volume} {5}},\ \bibinfo {pages}
  {034801} (\bibinfo {year} {2020}{\natexlab{b}})}\BibitemShut {NoStop}%
\bibitem [{\citenamefont {Yin}\ \emph {et~al.}(2021)\citenamefont {Yin},
  \citenamefont {Tu}, \citenamefont {Gong}, \citenamefont {Fu}, \citenamefont
  {Yan},\ and\ \citenamefont {Lei}}]{RbV3Sb5SC}%
  \BibitemOpen
  \bibfield  {author} {\bibinfo {author} {\bibfnamefont {Q.}~\bibnamefont
  {Yin}}, \bibinfo {author} {\bibfnamefont {Z.}~\bibnamefont {Tu}}, \bibinfo
  {author} {\bibfnamefont {C.}~\bibnamefont {Gong}}, \bibinfo {author}
  {\bibfnamefont {Y.}~\bibnamefont {Fu}}, \bibinfo {author} {\bibfnamefont
  {S.}~\bibnamefont {Yan}},\ and\ \bibinfo {author} {\bibfnamefont
  {H.}~\bibnamefont {Lei}},\ }\bibfield  {title} {\bibinfo {title}
  {{Superconductivity and normal-state properties of kagome metal RbV$_3$Sb$_5$
  single crystals}},\ }\href@noop {} {\bibfield  {journal} {\bibinfo  {journal}
  {Chin. Phys. Lett.}\ }\textbf {\bibinfo {volume} {38}},\ \bibinfo {pages}
  {037403} (\bibinfo {year} {2021})}\BibitemShut {NoStop}%
\bibitem [{\citenamefont {Peng}\ \emph {et~al.}(2021)\citenamefont {Peng},
  \citenamefont {Han}, \citenamefont {Pokharel}, \citenamefont {Shen},
  \citenamefont {Li}, \citenamefont {Hashimoto}, \citenamefont {Lu},
  \citenamefont {Ortiz}, \citenamefont {Luo}, \citenamefont {Li}, \citenamefont
  {Guo}, \citenamefont {Wang}, \citenamefont {Cui}, \citenamefont {Sun},
  \citenamefont {Qiao}, \citenamefont {Wilson},\ and\ \citenamefont
  {He}}]{PhysRevLett.127.266401}%
  \BibitemOpen
  \bibfield  {author} {\bibinfo {author} {\bibfnamefont {S.}~\bibnamefont
  {Peng}}, \bibinfo {author} {\bibfnamefont {Y.}~\bibnamefont {Han}}, \bibinfo
  {author} {\bibfnamefont {G.}~\bibnamefont {Pokharel}}, \bibinfo {author}
  {\bibfnamefont {J.}~\bibnamefont {Shen}}, \bibinfo {author} {\bibfnamefont
  {Z.}~\bibnamefont {Li}}, \bibinfo {author} {\bibfnamefont {M.}~\bibnamefont
  {Hashimoto}}, \bibinfo {author} {\bibfnamefont {D.}~\bibnamefont {Lu}},
  \bibinfo {author} {\bibfnamefont {B.~R.}\ \bibnamefont {Ortiz}}, \bibinfo
  {author} {\bibfnamefont {Y.}~\bibnamefont {Luo}}, \bibinfo {author}
  {\bibfnamefont {H.}~\bibnamefont {Li}}, \bibinfo {author} {\bibfnamefont
  {M.}~\bibnamefont {Guo}}, \bibinfo {author} {\bibfnamefont {B.}~\bibnamefont
  {Wang}}, \bibinfo {author} {\bibfnamefont {S.}~\bibnamefont {Cui}}, \bibinfo
  {author} {\bibfnamefont {Z.}~\bibnamefont {Sun}}, \bibinfo {author}
  {\bibfnamefont {Z.}~\bibnamefont {Qiao}}, \bibinfo {author} {\bibfnamefont
  {S.~D.}\ \bibnamefont {Wilson}},\ and\ \bibinfo {author} {\bibfnamefont
  {J.}~\bibnamefont {He}},\ }\bibfield  {title} {\bibinfo {title} {Realizing
  kagome band structure in two-dimensional kagome surface states of
  \text{$R{\mathrm{V}}_{6}{\mathrm{Sn}}_{6}$ ($R=\mathrm{Gd}$, Ho)}},\ }\href
  {https://doi.org/10.1103/PhysRevLett.127.266401} {\bibfield  {journal}
  {\bibinfo  {journal} {Phys. Rev. Lett.}\ }\textbf {\bibinfo {volume} {127}},\
  \bibinfo {pages} {266401} (\bibinfo {year} {2021})}\BibitemShut {NoStop}%
\bibitem [{\citenamefont {Wang}\ \emph {et~al.}(2021)\citenamefont {Wang},
  \citenamefont {Neubauer}, \citenamefont {Duan}, \citenamefont {Yin},
  \citenamefont {Fujitsu}, \citenamefont {Hosono}, \citenamefont {Ye},
  \citenamefont {Zhang}, \citenamefont {Chi}, \citenamefont {Krycka},
  \citenamefont {Lei},\ and\ \citenamefont {Dai}}]{PhysRevB.103.014416}%
  \BibitemOpen
  \bibfield  {author} {\bibinfo {author} {\bibfnamefont {Q.}~\bibnamefont
  {Wang}}, \bibinfo {author} {\bibfnamefont {K.~J.}\ \bibnamefont {Neubauer}},
  \bibinfo {author} {\bibfnamefont {C.}~\bibnamefont {Duan}}, \bibinfo {author}
  {\bibfnamefont {Q.}~\bibnamefont {Yin}}, \bibinfo {author} {\bibfnamefont
  {S.}~\bibnamefont {Fujitsu}}, \bibinfo {author} {\bibfnamefont
  {H.}~\bibnamefont {Hosono}}, \bibinfo {author} {\bibfnamefont
  {F.}~\bibnamefont {Ye}}, \bibinfo {author} {\bibfnamefont {R.}~\bibnamefont
  {Zhang}}, \bibinfo {author} {\bibfnamefont {S.}~\bibnamefont {Chi}}, \bibinfo
  {author} {\bibfnamefont {K.}~\bibnamefont {Krycka}}, \bibinfo {author}
  {\bibfnamefont {H.}~\bibnamefont {Lei}},\ and\ \bibinfo {author}
  {\bibfnamefont {P.}~\bibnamefont {Dai}},\ }\bibfield  {title} {\bibinfo
  {title} {{Field-induced topological Hall effect and double-fan spin structure
  with a $c$-axis component in the metallic kagome antiferromagnetic compound
  YMn$_6$Sn$_6$}},\ }\href {https://doi.org/10.1103/PhysRevB.103.014416}
  {\bibfield  {journal} {\bibinfo  {journal} {Phys. Rev. B}\ }\textbf {\bibinfo
  {volume} {103}},\ \bibinfo {pages} {014416} (\bibinfo {year}
  {2021})}\BibitemShut {NoStop}%
\bibitem [{\citenamefont {Pokharel}\ \emph {et~al.}(2021)\citenamefont
  {Pokharel}, \citenamefont {Teicher}, \citenamefont {Ortiz}, \citenamefont
  {Sarte}, \citenamefont {Wu}, \citenamefont {Peng}, \citenamefont {He},
  \citenamefont {Seshadri},\ and\ \citenamefont
  {Wilson}}]{PhysRevB.104.235139}%
  \BibitemOpen
  \bibfield  {author} {\bibinfo {author} {\bibfnamefont {G.}~\bibnamefont
  {Pokharel}}, \bibinfo {author} {\bibfnamefont {S.~M.~L.}\ \bibnamefont
  {Teicher}}, \bibinfo {author} {\bibfnamefont {B.~R.}\ \bibnamefont {Ortiz}},
  \bibinfo {author} {\bibfnamefont {P.~M.}\ \bibnamefont {Sarte}}, \bibinfo
  {author} {\bibfnamefont {G.}~\bibnamefont {Wu}}, \bibinfo {author}
  {\bibfnamefont {S.}~\bibnamefont {Peng}}, \bibinfo {author} {\bibfnamefont
  {J.}~\bibnamefont {He}}, \bibinfo {author} {\bibfnamefont {R.}~\bibnamefont
  {Seshadri}},\ and\ \bibinfo {author} {\bibfnamefont {S.~D.}\ \bibnamefont
  {Wilson}},\ }\bibfield  {title} {\bibinfo {title} {{Electronic properties of
  the topological kagome metals YV$_6$Sn$_6$ and GdV$_6$Sn$_6$}},\ }\href
  {https://doi.org/10.1103/PhysRevB.104.235139} {\bibfield  {journal} {\bibinfo
   {journal} {Phys. Rev. B}\ }\textbf {\bibinfo {volume} {104}},\ \bibinfo
  {pages} {235139} (\bibinfo {year} {2021})}\BibitemShut {NoStop}%
\bibitem [{\citenamefont {Pokharel}\ \emph {et~al.}(2022)\citenamefont
  {Pokharel}, \citenamefont {Ortiz}, \citenamefont {Chamorro}, \citenamefont
  {Sarte}, \citenamefont {Kautzsch}, \citenamefont {Wu}, \citenamefont {Ruff},\
  and\ \citenamefont {Wilson}}]{PhysRevMaterials104202}%
  \BibitemOpen
  \bibfield  {author} {\bibinfo {author} {\bibfnamefont {G.}~\bibnamefont
  {Pokharel}}, \bibinfo {author} {\bibfnamefont {B.}~\bibnamefont {Ortiz}},
  \bibinfo {author} {\bibfnamefont {J.}~\bibnamefont {Chamorro}}, \bibinfo
  {author} {\bibfnamefont {P.}~\bibnamefont {Sarte}}, \bibinfo {author}
  {\bibfnamefont {L.}~\bibnamefont {Kautzsch}}, \bibinfo {author}
  {\bibfnamefont {G.}~\bibnamefont {Wu}}, \bibinfo {author} {\bibfnamefont
  {J.}~\bibnamefont {Ruff}},\ and\ \bibinfo {author} {\bibfnamefont {S.~D.}\
  \bibnamefont {Wilson}},\ }\bibfield  {title} {\bibinfo {title} {{Highly
  anisotropic magnetism in the vanadium-based kagome metal TbV$_6$Sn$_6$}},\
  }\href {https://doi.org/10.1103/PhysRevMaterials.6.104202} {\bibfield
  {journal} {\bibinfo  {journal} {Phys. Rev. Mater.}\ }\textbf {\bibinfo
  {volume} {6}},\ \bibinfo {pages} {104202} (\bibinfo {year}
  {2022})}\BibitemShut {NoStop}%
\bibitem [{\citenamefont {Rosenberg}\ \emph {et~al.}(2022)\citenamefont
  {Rosenberg}, \citenamefont {DeStefano}, \citenamefont {Guo}, \citenamefont
  {Oh}, \citenamefont {Hashimoto}, \citenamefont {Lu}, \citenamefont
  {Birgeneau}, \citenamefont {Lee}, \citenamefont {Ke}, \citenamefont {Yi},\
  and\ \citenamefont {Chu}}]{PhysRevB.106.115139}%
  \BibitemOpen
  \bibfield  {author} {\bibinfo {author} {\bibfnamefont {E.}~\bibnamefont
  {Rosenberg}}, \bibinfo {author} {\bibfnamefont {J.~M.}\ \bibnamefont
  {DeStefano}}, \bibinfo {author} {\bibfnamefont {Y.}~\bibnamefont {Guo}},
  \bibinfo {author} {\bibfnamefont {J.~S.}\ \bibnamefont {Oh}}, \bibinfo
  {author} {\bibfnamefont {M.}~\bibnamefont {Hashimoto}}, \bibinfo {author}
  {\bibfnamefont {D.}~\bibnamefont {Lu}}, \bibinfo {author} {\bibfnamefont
  {R.~J.}\ \bibnamefont {Birgeneau}}, \bibinfo {author} {\bibfnamefont
  {Y.}~\bibnamefont {Lee}}, \bibinfo {author} {\bibfnamefont {L.}~\bibnamefont
  {Ke}}, \bibinfo {author} {\bibfnamefont {M.}~\bibnamefont {Yi}},\ and\
  \bibinfo {author} {\bibfnamefont {J.-H.}\ \bibnamefont {Chu}},\ }\bibfield
  {title} {\bibinfo {title} {{Uniaxial ferromagnetism in the kagome metal
  TbV$_6$Sn$_6$}},\ }\href {https://doi.org/10.1103/PhysRevB.106.115139}
  {\bibfield  {journal} {\bibinfo  {journal} {Phys. Rev. B}\ }\textbf {\bibinfo
  {volume} {106}},\ \bibinfo {pages} {115139} (\bibinfo {year}
  {2022})}\BibitemShut {NoStop}%
\bibitem [{\citenamefont {Ghimire}\ \emph {et~al.}(2020)\citenamefont
  {Ghimire}, \citenamefont {Dally}, \citenamefont {Poudel}, \citenamefont
  {Jones}, \citenamefont {Michel}, \citenamefont {Magar}, \citenamefont
  {Bleuel}, \citenamefont {McGuire}, \citenamefont {Jiang}, \citenamefont
  {Mitchell}, \citenamefont {Lynn},\ and\ \citenamefont
  {Mazin}}]{sciadv_abe2680}%
  \BibitemOpen
  \bibfield  {author} {\bibinfo {author} {\bibfnamefont {N.~J.}\ \bibnamefont
  {Ghimire}}, \bibinfo {author} {\bibfnamefont {R.~L.}\ \bibnamefont {Dally}},
  \bibinfo {author} {\bibfnamefont {L.}~\bibnamefont {Poudel}}, \bibinfo
  {author} {\bibfnamefont {D.~C.}\ \bibnamefont {Jones}}, \bibinfo {author}
  {\bibfnamefont {D.}~\bibnamefont {Michel}}, \bibinfo {author} {\bibfnamefont
  {N.~T.}\ \bibnamefont {Magar}}, \bibinfo {author} {\bibfnamefont
  {M.}~\bibnamefont {Bleuel}}, \bibinfo {author} {\bibfnamefont {M.~A.}\
  \bibnamefont {McGuire}}, \bibinfo {author} {\bibfnamefont {J.~S.}\
  \bibnamefont {Jiang}}, \bibinfo {author} {\bibfnamefont {J.~F.}\ \bibnamefont
  {Mitchell}}, \bibinfo {author} {\bibfnamefont {J.~W.}\ \bibnamefont {Lynn}},\
  and\ \bibinfo {author} {\bibfnamefont {I.~I.}\ \bibnamefont {Mazin}},\
  }\bibfield  {title} {\bibinfo {title} {Competing magnetic phases and
  fluctuation-driven scalar spin chirality in the kagome metal
  \text{YMn$_6$Sn$_6$}},\ }\href {https://doi.org/10.1126/sciadv.abe2680}
  {\bibfield  {journal} {\bibinfo  {journal} {Science Advances}\ }\textbf
  {\bibinfo {volume} {6}},\ \bibinfo {pages} {eabe2680} (\bibinfo {year}
  {2020})}\BibitemShut {NoStop}%
\bibitem [{\citenamefont {Arachchige}\ \emph {et~al.}(2022)\citenamefont
  {Arachchige}, \citenamefont {Meier}, \citenamefont {Marshall}, \citenamefont
  {Matsuoka}, \citenamefont {Xue}, \citenamefont {McGuire}, \citenamefont
  {Hermann}, \citenamefont {Cao},\ and\ \citenamefont
  {Mandrus}}]{PhysRevLett.129.216402}%
  \BibitemOpen
  \bibfield  {author} {\bibinfo {author} {\bibfnamefont {H.~W.~S.}\
  \bibnamefont {Arachchige}}, \bibinfo {author} {\bibfnamefont {W.~R.}\
  \bibnamefont {Meier}}, \bibinfo {author} {\bibfnamefont {M.}~\bibnamefont
  {Marshall}}, \bibinfo {author} {\bibfnamefont {T.}~\bibnamefont {Matsuoka}},
  \bibinfo {author} {\bibfnamefont {R.}~\bibnamefont {Xue}}, \bibinfo {author}
  {\bibfnamefont {M.~A.}\ \bibnamefont {McGuire}}, \bibinfo {author}
  {\bibfnamefont {R.~P.}\ \bibnamefont {Hermann}}, \bibinfo {author}
  {\bibfnamefont {H.}~\bibnamefont {Cao}},\ and\ \bibinfo {author}
  {\bibfnamefont {D.}~\bibnamefont {Mandrus}},\ }\bibfield  {title} {\bibinfo
  {title} {Charge density wave in kagome lattice intermetallic
  \text{${\mathrm{ScV}}_{6}{\mathrm{Sn}}_{6}$}},\ }\href
  {https://doi.org/10.1103/PhysRevLett.129.216402} {\bibfield  {journal}
  {\bibinfo  {journal} {Phys. Rev. Lett.}\ }\textbf {\bibinfo {volume} {129}},\
  \bibinfo {pages} {216402} (\bibinfo {year} {2022})}\BibitemShut {NoStop}%
\bibitem [{\citenamefont {Yin}\ \emph {et~al.}(2020)\citenamefont {Yin},
  \citenamefont {Ma}, \citenamefont {Cochran}, \citenamefont {Xu},
  \citenamefont {Zhang}, \citenamefont {Tien}, \citenamefont {Shumiya},
  \citenamefont {Cheng}, \citenamefont {Jiang}, \citenamefont {Lian},
  \citenamefont {Song}, \citenamefont {Chang}, \citenamefont {Belopolski},
  \citenamefont {Multer}, \citenamefont {Litskevich}, \citenamefont {Cheng},
  \citenamefont {Yang}, \citenamefont {Swidler}, \citenamefont {Zhou},
  \citenamefont {Lin}, \citenamefont {Neupert}, \citenamefont {Wang},
  \citenamefont {Yao}, \citenamefont {Chang}, \citenamefont {Jia},\ and\
  \citenamefont {Zahid~Hasan}}]{Yin_2020}%
  \BibitemOpen
  \bibfield  {author} {\bibinfo {author} {\bibfnamefont {J.-X.}\ \bibnamefont
  {Yin}}, \bibinfo {author} {\bibfnamefont {W.}~\bibnamefont {Ma}}, \bibinfo
  {author} {\bibfnamefont {T.~A.}\ \bibnamefont {Cochran}}, \bibinfo {author}
  {\bibfnamefont {X.}~\bibnamefont {Xu}}, \bibinfo {author} {\bibfnamefont
  {S.~S.}\ \bibnamefont {Zhang}}, \bibinfo {author} {\bibfnamefont {H.-J.}\
  \bibnamefont {Tien}}, \bibinfo {author} {\bibfnamefont {N.}~\bibnamefont
  {Shumiya}}, \bibinfo {author} {\bibfnamefont {G.}~\bibnamefont {Cheng}},
  \bibinfo {author} {\bibfnamefont {K.}~\bibnamefont {Jiang}}, \bibinfo
  {author} {\bibfnamefont {B.}~\bibnamefont {Lian}}, \bibinfo {author}
  {\bibfnamefont {Z.}~\bibnamefont {Song}}, \bibinfo {author} {\bibfnamefont
  {G.}~\bibnamefont {Chang}}, \bibinfo {author} {\bibfnamefont
  {I.}~\bibnamefont {Belopolski}}, \bibinfo {author} {\bibfnamefont
  {D.}~\bibnamefont {Multer}}, \bibinfo {author} {\bibfnamefont
  {M.}~\bibnamefont {Litskevich}}, \bibinfo {author} {\bibfnamefont {Z.-J.}\
  \bibnamefont {Cheng}}, \bibinfo {author} {\bibfnamefont {X.~P.}\ \bibnamefont
  {Yang}}, \bibinfo {author} {\bibfnamefont {B.}~\bibnamefont {Swidler}},
  \bibinfo {author} {\bibfnamefont {H.}~\bibnamefont {Zhou}}, \bibinfo {author}
  {\bibfnamefont {H.}~\bibnamefont {Lin}}, \bibinfo {author} {\bibfnamefont
  {T.}~\bibnamefont {Neupert}}, \bibinfo {author} {\bibfnamefont
  {Z.}~\bibnamefont {Wang}}, \bibinfo {author} {\bibfnamefont {N.}~\bibnamefont
  {Yao}}, \bibinfo {author} {\bibfnamefont {T.-R.}\ \bibnamefont {Chang}},
  \bibinfo {author} {\bibfnamefont {S.}~\bibnamefont {Jia}},\ and\ \bibinfo
  {author} {\bibfnamefont {M.}~\bibnamefont {Zahid~Hasan}},\ }\bibfield
  {title} {\bibinfo {title} {Quantum-limit chern topological magnetism in
  \text{TbMn$_6$Sn$_6$}},\ }\href {https://doi.org/10.1038/s41586-020-2482-7}
  {\bibfield  {journal} {\bibinfo  {journal} {Nature}\ }\textbf {\bibinfo
  {volume} {583}},\ \bibinfo {pages} {533} (\bibinfo {year}
  {2020})}\BibitemShut {NoStop}%
\bibitem [{\citenamefont {Zhang}\ \emph
  {et~al.}(2022{\natexlab{a}})\citenamefont {Zhang}, \citenamefont {Liu},
  \citenamefont {Cui}, \citenamefont {Guo}, \citenamefont {Wang}, \citenamefont
  {Shi}, \citenamefont {Zhang}, \citenamefont {Wang}, \citenamefont {Dong},
  \citenamefont {Sun}, \citenamefont {Dun},\ and\ \citenamefont
  {Cheng}}]{PhysRevMaterials.6.105001}%
  \BibitemOpen
  \bibfield  {author} {\bibinfo {author} {\bibfnamefont {X.}~\bibnamefont
  {Zhang}}, \bibinfo {author} {\bibfnamefont {Z.}~\bibnamefont {Liu}}, \bibinfo
  {author} {\bibfnamefont {Q.}~\bibnamefont {Cui}}, \bibinfo {author}
  {\bibfnamefont {Q.}~\bibnamefont {Guo}}, \bibinfo {author} {\bibfnamefont
  {N.}~\bibnamefont {Wang}}, \bibinfo {author} {\bibfnamefont {L.}~\bibnamefont
  {Shi}}, \bibinfo {author} {\bibfnamefont {H.}~\bibnamefont {Zhang}}, \bibinfo
  {author} {\bibfnamefont {W.}~\bibnamefont {Wang}}, \bibinfo {author}
  {\bibfnamefont {X.}~\bibnamefont {Dong}}, \bibinfo {author} {\bibfnamefont
  {J.}~\bibnamefont {Sun}}, \bibinfo {author} {\bibfnamefont {Z.}~\bibnamefont
  {Dun}},\ and\ \bibinfo {author} {\bibfnamefont {J.}~\bibnamefont {Cheng}},\
  }\bibfield  {title} {\bibinfo {title} {{Electronic and magnetic properties of
  intermetallic kagome magnets \textit{R}V$_6$Sn$_6$ (\textit{R}: Tb--Tm)}},\
  }\href {https://doi.org/10.1103/PhysRevMaterials.6.105001} {\bibfield
  {journal} {\bibinfo  {journal} {Phys. Rev. Mater.}\ }\textbf {\bibinfo
  {volume} {6}},\ \bibinfo {pages} {105001} (\bibinfo {year}
  {2022}{\natexlab{a}})}\BibitemShut {NoStop}%
\bibitem [{\citenamefont {Lee}\ and\ \citenamefont
  {Mun}(2022)}]{PhysRevMaterials.6.083401}%
  \BibitemOpen
  \bibfield  {author} {\bibinfo {author} {\bibfnamefont {J.}~\bibnamefont
  {Lee}}\ and\ \bibinfo {author} {\bibfnamefont {E.}~\bibnamefont {Mun}},\
  }\bibfield  {title} {\bibinfo {title} {Anisotropic magnetic property of
  single crystals \text{$R{\mathrm{V}}_{6}{\mathrm{Sn}}_{6}$ $(R=\mathrm{Y},
  \mathrm{Gd}\text{\ensuremath{-}}\mathrm{Tm}, \mathrm{Lu})$}},\ }\href
  {https://doi.org/10.1103/PhysRevMaterials.6.083401} {\bibfield  {journal}
  {\bibinfo  {journal} {Phys. Rev. Mater.}\ }\textbf {\bibinfo {volume} {6}},\
  \bibinfo {pages} {083401} (\bibinfo {year} {2022})}\BibitemShut {NoStop}%
\bibitem [{\citenamefont {Shao-ying}\ \emph {et~al.}(2001)\citenamefont
  {Shao-ying}, \citenamefont {Peng}, \citenamefont {Run-wei}, \citenamefont
  {Sun Ji-rong}, \citenamefont {Hong-wei},\ and\ \citenamefont
  {Bao-gen}}]{ZhangShao-ying_2001}%
  \BibitemOpen
  \bibfield  {author} {\bibinfo {author} {\bibfnamefont {Z.}~\bibnamefont
  {Shao-ying}}, \bibinfo {author} {\bibfnamefont {Z.}~\bibnamefont {Peng}},
  \bibinfo {author} {\bibfnamefont {L.}~\bibnamefont {Run-wei}}, \bibinfo
  {author} {\bibfnamefont {C.~Z.-h.}\ \bibnamefont {Sun Ji-rong}}, \bibinfo
  {author} {\bibfnamefont {Z.}~\bibnamefont {Hong-wei}},\ and\ \bibinfo
  {author} {\bibfnamefont {S.}~\bibnamefont {Bao-gen}},\ }\bibfield  {title}
  {\bibinfo {title} {Structure, magnetic properties and giant magnetoresistance
  of \text{YMn$_6$Sn$_{6-x}$Ga$_x$ (x = 0-0.6)} compounds},\ }\href
  {https://doi.org/10.1088/1009-1963/10/4/318} {\bibfield  {journal} {\bibinfo
  {journal} {Chin. Phys.}\ }\textbf {\bibinfo {volume} {10}},\ \bibinfo {pages}
  {345} (\bibinfo {year} {2001})}\BibitemShut {NoStop}%
\bibitem [{\citenamefont {Ma}\ \emph {et~al.}(2021)\citenamefont {Ma},
  \citenamefont {Xu}, \citenamefont {Yin}, \citenamefont {Yang}, \citenamefont
  {Zhou}, \citenamefont {Cheng}, \citenamefont {Huang}, \citenamefont {Qu},
  \citenamefont {Wang}, \citenamefont {Hasan},\ and\ \citenamefont
  {Jia}}]{PhysRevLett.126.246602}%
  \BibitemOpen
  \bibfield  {author} {\bibinfo {author} {\bibfnamefont {W.}~\bibnamefont
  {Ma}}, \bibinfo {author} {\bibfnamefont {X.}~\bibnamefont {Xu}}, \bibinfo
  {author} {\bibfnamefont {J.-X.}\ \bibnamefont {Yin}}, \bibinfo {author}
  {\bibfnamefont {H.}~\bibnamefont {Yang}}, \bibinfo {author} {\bibfnamefont
  {H.}~\bibnamefont {Zhou}}, \bibinfo {author} {\bibfnamefont {Z.-J.}\
  \bibnamefont {Cheng}}, \bibinfo {author} {\bibfnamefont {Y.}~\bibnamefont
  {Huang}}, \bibinfo {author} {\bibfnamefont {Z.}~\bibnamefont {Qu}}, \bibinfo
  {author} {\bibfnamefont {F.}~\bibnamefont {Wang}}, \bibinfo {author}
  {\bibfnamefont {M.~Z.}\ \bibnamefont {Hasan}},\ and\ \bibinfo {author}
  {\bibfnamefont {S.}~\bibnamefont {Jia}},\ }\bibfield  {title} {\bibinfo
  {title} {Rare earth engineering in
  \text{$R{\mathrm{Mn}}_{6}{\mathrm{Sn}}_{6}$
  ($R=\text{Gd}\text{\ensuremath{-}}\text{Tm}$, Lu)} topological kagome
  magnets},\ }\href {https://doi.org/10.1103/PhysRevLett.126.246602} {\bibfield
   {journal} {\bibinfo  {journal} {Phys. Rev. Lett.}\ }\textbf {\bibinfo
  {volume} {126}},\ \bibinfo {pages} {246602} (\bibinfo {year}
  {2021})}\BibitemShut {NoStop}%
\bibitem [{\citenamefont {Park}\ \emph {et~al.}(2021)\citenamefont {Park},
  \citenamefont {Ye},\ and\ \citenamefont {Balents}}]{park2021electronic}%
  \BibitemOpen
  \bibfield  {author} {\bibinfo {author} {\bibfnamefont {T.}~\bibnamefont
  {Park}}, \bibinfo {author} {\bibfnamefont {M.}~\bibnamefont {Ye}},\ and\
  \bibinfo {author} {\bibfnamefont {L.}~\bibnamefont {Balents}},\ }\bibfield
  {title} {\bibinfo {title} {{Electronic instabilities of kagome metals: saddle
  points and Landau theory}},\ }\href@noop {} {\bibfield  {journal} {\bibinfo
  {journal} {Phys. Rev. B}\ }\textbf {\bibinfo {volume} {104}},\ \bibinfo
  {pages} {035142} (\bibinfo {year} {2021})}\BibitemShut {NoStop}%
\bibitem [{\citenamefont {Wang}\ \emph {et~al.}(2013)\citenamefont {Wang},
  \citenamefont {Li}, \citenamefont {Xiang},\ and\ \citenamefont
  {Wang}}]{PhysRevB.87.115135}%
  \BibitemOpen
  \bibfield  {author} {\bibinfo {author} {\bibfnamefont {W.-S.}\ \bibnamefont
  {Wang}}, \bibinfo {author} {\bibfnamefont {Z.-Z.}\ \bibnamefont {Li}},
  \bibinfo {author} {\bibfnamefont {Y.-Y.}\ \bibnamefont {Xiang}},\ and\
  \bibinfo {author} {\bibfnamefont {Q.-H.}\ \bibnamefont {Wang}},\ }\bibfield
  {title} {\bibinfo {title} {Competing electronic orders on kagome lattices at
  van hove filling},\ }\href {https://doi.org/10.1103/PhysRevB.87.115135}
  {\bibfield  {journal} {\bibinfo  {journal} {Phys. Rev. B}\ }\textbf {\bibinfo
  {volume} {87}},\ \bibinfo {pages} {115135} (\bibinfo {year}
  {2013})}\BibitemShut {NoStop}%
\bibitem [{\citenamefont {Kiesel}\ \emph {et~al.}(2013)\citenamefont {Kiesel},
  \citenamefont {Platt},\ and\ \citenamefont
  {Thomale}}]{kiesel2013unconventional}%
  \BibitemOpen
  \bibfield  {author} {\bibinfo {author} {\bibfnamefont {M.~L.}\ \bibnamefont
  {Kiesel}}, \bibinfo {author} {\bibfnamefont {C.}~\bibnamefont {Platt}},\ and\
  \bibinfo {author} {\bibfnamefont {R.}~\bibnamefont {Thomale}},\ }\bibfield
  {title} {\bibinfo {title} {{Unconventional Fermi surface instabilities in the
  kagome Hubbard model}},\ }\href@noop {} {\bibfield  {journal} {\bibinfo
  {journal} {Phys. Rev. Lett.}\ }\textbf {\bibinfo {volume} {110}},\ \bibinfo
  {pages} {126405} (\bibinfo {year} {2013})}\BibitemShut {NoStop}%
\bibitem [{\citenamefont {Meier}\ \emph {et~al.}(2020)\citenamefont {Meier},
  \citenamefont {Du}, \citenamefont {Okamoto}, \citenamefont {Mohanta},
  \citenamefont {May}, \citenamefont {McGuire}, \citenamefont {Bridges},
  \citenamefont {Samolyuk},\ and\ \citenamefont {Sales}}]{meier2020flat}%
  \BibitemOpen
  \bibfield  {author} {\bibinfo {author} {\bibfnamefont {W.~R.}\ \bibnamefont
  {Meier}}, \bibinfo {author} {\bibfnamefont {M.-H.}\ \bibnamefont {Du}},
  \bibinfo {author} {\bibfnamefont {S.}~\bibnamefont {Okamoto}}, \bibinfo
  {author} {\bibfnamefont {N.}~\bibnamefont {Mohanta}}, \bibinfo {author}
  {\bibfnamefont {A.~F.}\ \bibnamefont {May}}, \bibinfo {author} {\bibfnamefont
  {M.~A.}\ \bibnamefont {McGuire}}, \bibinfo {author} {\bibfnamefont {C.~A.}\
  \bibnamefont {Bridges}}, \bibinfo {author} {\bibfnamefont {G.~D.}\
  \bibnamefont {Samolyuk}},\ and\ \bibinfo {author} {\bibfnamefont {B.~C.}\
  \bibnamefont {Sales}},\ }\bibfield  {title} {\bibinfo {title} {{Flat bands in
  the CoSn-type compounds}},\ }\href@noop {} {\bibfield  {journal} {\bibinfo
  {journal} {Phys. Rev. B}\ }\textbf {\bibinfo {volume} {102}},\ \bibinfo
  {pages} {075148} (\bibinfo {year} {2020})}\BibitemShut {NoStop}%
\bibitem [{\citenamefont {Isakov}\ \emph {et~al.}(2006)\citenamefont {Isakov},
  \citenamefont {Wessel}, \citenamefont {Melko}, \citenamefont {Sengupta},\
  and\ \citenamefont {Kim}}]{PhysRevLett.97.147202}%
  \BibitemOpen
  \bibfield  {author} {\bibinfo {author} {\bibfnamefont {S.~V.}\ \bibnamefont
  {Isakov}}, \bibinfo {author} {\bibfnamefont {S.}~\bibnamefont {Wessel}},
  \bibinfo {author} {\bibfnamefont {R.~G.}\ \bibnamefont {Melko}}, \bibinfo
  {author} {\bibfnamefont {K.}~\bibnamefont {Sengupta}},\ and\ \bibinfo
  {author} {\bibfnamefont {Y.~B.}\ \bibnamefont {Kim}},\ }\bibfield  {title}
  {\bibinfo {title} {Hard-core bosons on the kagome lattice: Valence-bond
  solids and their quantum melting},\ }\href
  {https://doi.org/10.1103/PhysRevLett.97.147202} {\bibfield  {journal}
  {\bibinfo  {journal} {Phys. Rev. Lett.}\ }\textbf {\bibinfo {volume} {97}},\
  \bibinfo {pages} {147202} (\bibinfo {year} {2006})}\BibitemShut {NoStop}%
\bibitem [{\citenamefont {O'Brien}\ \emph {et~al.}(2010)\citenamefont
  {O'Brien}, \citenamefont {Pollmann},\ and\ \citenamefont
  {Fulde}}]{PhysRevB.81.235115}%
  \BibitemOpen
  \bibfield  {author} {\bibinfo {author} {\bibfnamefont {A.}~\bibnamefont
  {O'Brien}}, \bibinfo {author} {\bibfnamefont {F.}~\bibnamefont {Pollmann}},\
  and\ \bibinfo {author} {\bibfnamefont {P.}~\bibnamefont {Fulde}},\ }\bibfield
   {title} {\bibinfo {title} {Strongly correlated fermions on a kagome
  lattice},\ }\href {https://doi.org/10.1103/PhysRevB.81.235115} {\bibfield
  {journal} {\bibinfo  {journal} {Phys. Rev. B}\ }\textbf {\bibinfo {volume}
  {81}},\ \bibinfo {pages} {235115} (\bibinfo {year} {2010})}\BibitemShut
  {NoStop}%
\bibitem [{\citenamefont {R\"{u}egg}\ and\ \citenamefont
  {Fiete}(2011)}]{PhysRevB.83.165118}%
  \BibitemOpen
  \bibfield  {author} {\bibinfo {author} {\bibfnamefont {A.}~\bibnamefont
  {R\"{u}egg}}\ and\ \bibinfo {author} {\bibfnamefont {G.~A.}\ \bibnamefont
  {Fiete}},\ }\bibfield  {title} {\bibinfo {title} {Fractionally charged
  topological point defects on the kagome lattice},\ }\href
  {https://doi.org/10.1103/PhysRevB.83.165118} {\bibfield  {journal} {\bibinfo
  {journal} {Phys. Rev. B}\ }\textbf {\bibinfo {volume} {83}},\ \bibinfo
  {pages} {165118} (\bibinfo {year} {2011})}\BibitemShut {NoStop}%
\bibitem [{\citenamefont {Guo}\ and\ \citenamefont
  {Franz}(2009)}]{PhysRevB.80.113102}%
  \BibitemOpen
  \bibfield  {author} {\bibinfo {author} {\bibfnamefont {H.-M.}\ \bibnamefont
  {Guo}}\ and\ \bibinfo {author} {\bibfnamefont {M.}~\bibnamefont {Franz}},\
  }\bibfield  {title} {\bibinfo {title} {Topological insulator on the kagome
  lattice},\ }\href {https://doi.org/10.1103/PhysRevB.80.113102} {\bibfield
  {journal} {\bibinfo  {journal} {Phys. Rev. B}\ }\textbf {\bibinfo {volume}
  {80}},\ \bibinfo {pages} {113102} (\bibinfo {year} {2009})}\BibitemShut
  {NoStop}%
\bibitem [{\citenamefont {Yu}\ and\ \citenamefont {Li}(2012)}]{yu2012chiral}%
  \BibitemOpen
  \bibfield  {author} {\bibinfo {author} {\bibfnamefont {S.-L.}\ \bibnamefont
  {Yu}}\ and\ \bibinfo {author} {\bibfnamefont {J.-X.}\ \bibnamefont {Li}},\
  }\bibfield  {title} {\bibinfo {title} {{Chiral superconducting phase and
  chiral spin-density-wave phase in a Hubbard model on the kagome lattice}},\
  }\href@noop {} {\bibfield  {journal} {\bibinfo  {journal} {Phys. Rev. B}\
  }\textbf {\bibinfo {volume} {85}},\ \bibinfo {pages} {144402} (\bibinfo
  {year} {2012})}\BibitemShut {NoStop}%
\bibitem [{\citenamefont {Ko}\ \emph {et~al.}(2009)\citenamefont {Ko},
  \citenamefont {Lee},\ and\ \citenamefont {Wen}}]{ko2009doped}%
  \BibitemOpen
  \bibfield  {author} {\bibinfo {author} {\bibfnamefont {W.-H.}\ \bibnamefont
  {Ko}}, \bibinfo {author} {\bibfnamefont {P.~A.}\ \bibnamefont {Lee}},\ and\
  \bibinfo {author} {\bibfnamefont {X.-G.}\ \bibnamefont {Wen}},\ }\bibfield
  {title} {\bibinfo {title} {{Doped kagome system as exotic superconductor}},\
  }\href@noop {} {\bibfield  {journal} {\bibinfo  {journal} {Phys. Rev. B}\
  }\textbf {\bibinfo {volume} {79}},\ \bibinfo {pages} {214502} (\bibinfo
  {year} {2009})}\BibitemShut {NoStop}%
\bibitem [{\citenamefont {Ortiz}\ \emph {et~al.}(2021)\citenamefont {Ortiz},
  \citenamefont {Teicher}, \citenamefont {Kautzsch}, \citenamefont {Sarte},
  \citenamefont {Ratcliff}, \citenamefont {Harter}, \citenamefont {Ruff},
  \citenamefont {Seshadri},\ and\ \citenamefont {Wilson}}]{ortiz2021fermi}%
  \BibitemOpen
  \bibfield  {author} {\bibinfo {author} {\bibfnamefont {B.~R.}\ \bibnamefont
  {Ortiz}}, \bibinfo {author} {\bibfnamefont {S.~M.}\ \bibnamefont {Teicher}},
  \bibinfo {author} {\bibfnamefont {L.}~\bibnamefont {Kautzsch}}, \bibinfo
  {author} {\bibfnamefont {P.~M.}\ \bibnamefont {Sarte}}, \bibinfo {author}
  {\bibfnamefont {N.}~\bibnamefont {Ratcliff}}, \bibinfo {author}
  {\bibfnamefont {J.}~\bibnamefont {Harter}}, \bibinfo {author} {\bibfnamefont
  {J.~P.}\ \bibnamefont {Ruff}}, \bibinfo {author} {\bibfnamefont
  {R.}~\bibnamefont {Seshadri}},\ and\ \bibinfo {author} {\bibfnamefont
  {S.~D.}\ \bibnamefont {Wilson}},\ }\bibfield  {title} {\bibinfo {title}
  {{Fermi surface mapping and the nature of charge-density-wave order in the
  kagome superconductor CsV$_3$Sb$_5$}},\ }\href@noop {} {\bibfield  {journal}
  {\bibinfo  {journal} {Phys. Rev. X}\ }\textbf {\bibinfo {volume} {11}},\
  \bibinfo {pages} {041030} (\bibinfo {year} {2021})}\BibitemShut {NoStop}%
\bibitem [{\citenamefont {Zhao}\ \emph {et~al.}(2021)\citenamefont {Zhao},
  \citenamefont {Li}, \citenamefont {Ortiz}, \citenamefont {Teicher},
  \citenamefont {Park}, \citenamefont {Ye}, \citenamefont {Wang}, \citenamefont
  {Balents}, \citenamefont {Wilson},\ and\ \citenamefont
  {Zeljkovic}}]{zhao2021cascade}%
  \BibitemOpen
  \bibfield  {author} {\bibinfo {author} {\bibfnamefont {H.}~\bibnamefont
  {Zhao}}, \bibinfo {author} {\bibfnamefont {H.}~\bibnamefont {Li}}, \bibinfo
  {author} {\bibfnamefont {B.~R.}\ \bibnamefont {Ortiz}}, \bibinfo {author}
  {\bibfnamefont {S.~M.}\ \bibnamefont {Teicher}}, \bibinfo {author}
  {\bibfnamefont {T.}~\bibnamefont {Park}}, \bibinfo {author} {\bibfnamefont
  {M.}~\bibnamefont {Ye}}, \bibinfo {author} {\bibfnamefont {Z.}~\bibnamefont
  {Wang}}, \bibinfo {author} {\bibfnamefont {L.}~\bibnamefont {Balents}},
  \bibinfo {author} {\bibfnamefont {S.~D.}\ \bibnamefont {Wilson}},\ and\
  \bibinfo {author} {\bibfnamefont {I.}~\bibnamefont {Zeljkovic}},\ }\bibfield
  {title} {\bibinfo {title} {{Cascade of correlated electron states in the
  kagome superconductor CsV$_3$Sb$_5$}},\ }\href@noop {} {\bibfield  {journal}
  {\bibinfo  {journal} {Nature}\ }\textbf {\bibinfo {volume} {599}},\ \bibinfo
  {pages} {216} (\bibinfo {year} {2021})}\BibitemShut {NoStop}%
\bibitem [{\citenamefont {Hu}\ \emph {et~al.}(2022)\citenamefont {Hu},
  \citenamefont {Wu}, \citenamefont {Ortiz}, \citenamefont {Han}, \citenamefont
  {Plumb}, \citenamefont {Wilson}, \citenamefont {Schnyder}, \citenamefont
  {Shi} \emph {et~al.}}]{hu2022coexistence}%
  \BibitemOpen
  \bibfield  {author} {\bibinfo {author} {\bibfnamefont {Y.}~\bibnamefont
  {Hu}}, \bibinfo {author} {\bibfnamefont {X.}~\bibnamefont {Wu}}, \bibinfo
  {author} {\bibfnamefont {B.~R.}\ \bibnamefont {Ortiz}}, \bibinfo {author}
  {\bibfnamefont {X.}~\bibnamefont {Han}}, \bibinfo {author} {\bibfnamefont
  {N.~C.}\ \bibnamefont {Plumb}}, \bibinfo {author} {\bibfnamefont {S.~D.}\
  \bibnamefont {Wilson}}, \bibinfo {author} {\bibfnamefont {A.~P.}\
  \bibnamefont {Schnyder}}, \bibinfo {author} {\bibfnamefont {M.}~\bibnamefont
  {Shi}}, \emph {et~al.},\ }\bibfield  {title} {\bibinfo {title} {{Coexistence
  of trihexagonal and star-of-David pattern in the charge density wave of the
  kagome superconductor \textit{A}V$_3$Sb$_5$}},\ }\href@noop {} {\bibfield
  {journal} {\bibinfo  {journal} {Phys. Rev. B}\ }\textbf {\bibinfo {volume}
  {106}},\ \bibinfo {pages} {L241106} (\bibinfo {year} {2022})}\BibitemShut
  {NoStop}%
\bibitem [{\citenamefont {Kang}\ \emph {et~al.}(2023)\citenamefont {Kang},
  \citenamefont {Fang}, \citenamefont {Yoo}, \citenamefont {Ortiz},
  \citenamefont {Oey}, \citenamefont {Choi}, \citenamefont {Ryu}, \citenamefont
  {Kim}, \citenamefont {Jozwiak}, \citenamefont {Bostwick} \emph
  {et~al.}}]{kang2022microscopic}%
  \BibitemOpen
  \bibfield  {author} {\bibinfo {author} {\bibfnamefont {M.}~\bibnamefont
  {Kang}}, \bibinfo {author} {\bibfnamefont {S.}~\bibnamefont {Fang}}, \bibinfo
  {author} {\bibfnamefont {J.}~\bibnamefont {Yoo}}, \bibinfo {author}
  {\bibfnamefont {B.~R.}\ \bibnamefont {Ortiz}}, \bibinfo {author}
  {\bibfnamefont {Y.~M.}\ \bibnamefont {Oey}}, \bibinfo {author} {\bibfnamefont
  {J.}~\bibnamefont {Choi}}, \bibinfo {author} {\bibfnamefont {S.~H.}\
  \bibnamefont {Ryu}}, \bibinfo {author} {\bibfnamefont {J.}~\bibnamefont
  {Kim}}, \bibinfo {author} {\bibfnamefont {C.}~\bibnamefont {Jozwiak}},
  \bibinfo {author} {\bibfnamefont {A.}~\bibnamefont {Bostwick}}, \emph
  {et~al.},\ }\bibfield  {title} {\bibinfo {title} {Charge order landscape and
  competition with superconductivity in kagome metals},\ }\href@noop {}
  {\bibfield  {journal} {\bibinfo  {journal} {Nat. Mater.}\ }\textbf {\bibinfo
  {volume} {22}},\ \bibinfo {pages} {186} (\bibinfo {year} {2023})}\BibitemShut
  {NoStop}%
\bibitem [{\citenamefont {Jiang}\ \emph {et~al.}(2021)\citenamefont {Jiang},
  \citenamefont {Yin}, \citenamefont {Denner}, \citenamefont {Shumiya},
  \citenamefont {Ortiz}, \citenamefont {Xu}, \citenamefont {Guguchia},
  \citenamefont {He}, \citenamefont {Hossain}, \citenamefont {Liu} \emph
  {et~al.}}]{jiang2021unconventional}%
  \BibitemOpen
  \bibfield  {author} {\bibinfo {author} {\bibfnamefont {Y.-X.}\ \bibnamefont
  {Jiang}}, \bibinfo {author} {\bibfnamefont {J.-X.}\ \bibnamefont {Yin}},
  \bibinfo {author} {\bibfnamefont {M.~M.}\ \bibnamefont {Denner}}, \bibinfo
  {author} {\bibfnamefont {N.}~\bibnamefont {Shumiya}}, \bibinfo {author}
  {\bibfnamefont {B.~R.}\ \bibnamefont {Ortiz}}, \bibinfo {author}
  {\bibfnamefont {G.}~\bibnamefont {Xu}}, \bibinfo {author} {\bibfnamefont
  {Z.}~\bibnamefont {Guguchia}}, \bibinfo {author} {\bibfnamefont
  {J.}~\bibnamefont {He}}, \bibinfo {author} {\bibfnamefont {M.~S.}\
  \bibnamefont {Hossain}}, \bibinfo {author} {\bibfnamefont {X.}~\bibnamefont
  {Liu}}, \emph {et~al.},\ }\bibfield  {title} {\bibinfo {title}
  {{Unconventional chiral charge order in kagome superconductor
  KV$_3$Sb$_5$}},\ }\href@noop {} {\bibfield  {journal} {\bibinfo  {journal}
  {Nat. Mater.}\ }\textbf {\bibinfo {volume} {20}},\ \bibinfo {pages} {1353}
  (\bibinfo {year} {2021})}\BibitemShut {NoStop}%
\bibitem [{\citenamefont {Zhang}\ \emph
  {et~al.}(2022{\natexlab{b}})\citenamefont {Zhang}, \citenamefont {Liu},
  \citenamefont {Zhang}, \citenamefont {Hou}, \citenamefont {Fu}, \citenamefont
  {Zhang}, \citenamefont {Gao},\ and\ \citenamefont {Liu}}]{zhang2022magnetic}%
  \BibitemOpen
  \bibfield  {author} {\bibinfo {author} {\bibfnamefont {H.}~\bibnamefont
  {Zhang}}, \bibinfo {author} {\bibfnamefont {C.}~\bibnamefont {Liu}}, \bibinfo
  {author} {\bibfnamefont {Y.}~\bibnamefont {Zhang}}, \bibinfo {author}
  {\bibfnamefont {Z.}~\bibnamefont {Hou}}, \bibinfo {author} {\bibfnamefont
  {X.}~\bibnamefont {Fu}}, \bibinfo {author} {\bibfnamefont {X.}~\bibnamefont
  {Zhang}}, \bibinfo {author} {\bibfnamefont {X.}~\bibnamefont {Gao}},\ and\
  \bibinfo {author} {\bibfnamefont {J.}~\bibnamefont {Liu}},\ }\bibfield
  {title} {\bibinfo {title} {{Magnetic field-induced nontrivial spin chirality
  and large topological Hall effect in kagome magnet ScMn$_6$Sn$_6$}},\
  }\href@noop {} {\bibfield  {journal} {\bibinfo  {journal} {Appl. Phys.
  Lett.}\ }\textbf {\bibinfo {volume} {121}},\ \bibinfo {pages} {202401}
  (\bibinfo {year} {2022}{\natexlab{b}})}\BibitemShut {NoStop}%
\bibitem [{\citenamefont {Dhakal}\ \emph {et~al.}(2021)\citenamefont {Dhakal},
  \citenamefont {Kabeer}, \citenamefont {Pathak}, \citenamefont {Kabir},
  \citenamefont {Poudel}, \citenamefont {Filippone}, \citenamefont {Casey},
  \citenamefont {Sakhya}, \citenamefont {Regmi}, \citenamefont {Sims} \emph
  {et~al.}}]{dhakal2021anisotropically}%
  \BibitemOpen
  \bibfield  {author} {\bibinfo {author} {\bibfnamefont {G.}~\bibnamefont
  {Dhakal}}, \bibinfo {author} {\bibfnamefont {F.~C.}\ \bibnamefont {Kabeer}},
  \bibinfo {author} {\bibfnamefont {A.~K.}\ \bibnamefont {Pathak}}, \bibinfo
  {author} {\bibfnamefont {F.}~\bibnamefont {Kabir}}, \bibinfo {author}
  {\bibfnamefont {N.}~\bibnamefont {Poudel}}, \bibinfo {author} {\bibfnamefont
  {R.}~\bibnamefont {Filippone}}, \bibinfo {author} {\bibfnamefont
  {J.}~\bibnamefont {Casey}}, \bibinfo {author} {\bibfnamefont {A.~P.}\
  \bibnamefont {Sakhya}}, \bibinfo {author} {\bibfnamefont {S.}~\bibnamefont
  {Regmi}}, \bibinfo {author} {\bibfnamefont {C.}~\bibnamefont {Sims}}, \emph
  {et~al.},\ }\bibfield  {title} {\bibinfo {title} {Anisotropically large
  anomalous and topological hall effect in a kagome magnet},\ }\href@noop {}
  {\bibfield  {journal} {\bibinfo  {journal} {Physical Review B}\ }\textbf
  {\bibinfo {volume} {104}},\ \bibinfo {pages} {L161115} (\bibinfo {year}
  {2021})}\BibitemShut {NoStop}%
\bibitem [{\citenamefont {Teng}\ \emph {et~al.}(2022)\citenamefont {Teng},
  \citenamefont {Chen}, \citenamefont {Ye}, \citenamefont {Rosenberg},
  \citenamefont {Liu}, \citenamefont {Yin}, \citenamefont {Jiang},
  \citenamefont {Oh}, \citenamefont {Hasan}, \citenamefont {Neubauer} \emph
  {et~al.}}]{teng2022discovery}%
  \BibitemOpen
  \bibfield  {author} {\bibinfo {author} {\bibfnamefont {X.}~\bibnamefont
  {Teng}}, \bibinfo {author} {\bibfnamefont {L.}~\bibnamefont {Chen}}, \bibinfo
  {author} {\bibfnamefont {F.}~\bibnamefont {Ye}}, \bibinfo {author}
  {\bibfnamefont {E.}~\bibnamefont {Rosenberg}}, \bibinfo {author}
  {\bibfnamefont {Z.}~\bibnamefont {Liu}}, \bibinfo {author} {\bibfnamefont
  {J.-X.}\ \bibnamefont {Yin}}, \bibinfo {author} {\bibfnamefont {Y.-X.}\
  \bibnamefont {Jiang}}, \bibinfo {author} {\bibfnamefont {J.~S.}\ \bibnamefont
  {Oh}}, \bibinfo {author} {\bibfnamefont {M.~Z.}\ \bibnamefont {Hasan}},
  \bibinfo {author} {\bibfnamefont {K.~J.}\ \bibnamefont {Neubauer}}, \emph
  {et~al.},\ }\bibfield  {title} {\bibinfo {title} {Discovery of charge density
  wave in a kagome lattice antiferromagnet},\ }\href@noop {} {\bibfield
  {journal} {\bibinfo  {journal} {Nature}\ }\textbf {\bibinfo {volume} {609}},\
  \bibinfo {pages} {490} (\bibinfo {year} {2022})}\BibitemShut {NoStop}%
\bibitem [{\citenamefont {Ovchinnikov}\ and\ \citenamefont
  {Bobev}(2018)}]{ovchinnikov2018synthesis}%
  \BibitemOpen
  \bibfield  {author} {\bibinfo {author} {\bibfnamefont {A.}~\bibnamefont
  {Ovchinnikov}}\ and\ \bibinfo {author} {\bibfnamefont {S.}~\bibnamefont
  {Bobev}},\ }\bibfield  {title} {\bibinfo {title} {{Synthesis, Crystal and
  Electronic Structure of the Titanium Bismuthides Sr$_5$Ti$_{12}$Bi$_{19+x}$,
  Ba$_5$Ti$_{12}$Bi$_{19+x}$, and
  Sr$_{5-\delta}$Eu$_\delta$Ti$_{12}$Bi$_{19+x}$ (x=0.5--1.0; $\delta$=2.4,
  4.0)}},\ }\href@noop {} {\bibfield  {journal} {\bibinfo  {journal} {Eur. J.
  Inorg. Chem.}\ }\textbf {\bibinfo {volume} {2018}},\ \bibinfo {pages} {1266}
  (\bibinfo {year} {2018})}\BibitemShut {NoStop}%
\bibitem [{\citenamefont {Ovchinnikov}\ and\ \citenamefont
  {Bobev}(2019)}]{ovchinnikov2019bismuth}%
  \BibitemOpen
  \bibfield  {author} {\bibinfo {author} {\bibfnamefont {A.}~\bibnamefont
  {Ovchinnikov}}\ and\ \bibinfo {author} {\bibfnamefont {S.}~\bibnamefont
  {Bobev}},\ }\bibfield  {title} {\bibinfo {title} {Bismuth as a reactive
  solvent in the synthesis of multicomponent transition-metal-bearing
  bismuthides},\ }\href@noop {} {\bibfield  {journal} {\bibinfo  {journal}
  {Inorg. Chem.}\ }\textbf {\bibinfo {volume} {59}},\ \bibinfo {pages} {3459}
  (\bibinfo {year} {2019})}\BibitemShut {NoStop}%
\bibitem [{\citenamefont {Werhahn}\ \emph {et~al.}(2022)\citenamefont
  {Werhahn}, \citenamefont {Ortiz}, \citenamefont {Hay}, \citenamefont
  {Wilson}, \citenamefont {Seshadri},\ and\ \citenamefont
  {Johrendt}}]{werhahn2022kagome}%
  \BibitemOpen
  \bibfield  {author} {\bibinfo {author} {\bibfnamefont {D.}~\bibnamefont
  {Werhahn}}, \bibinfo {author} {\bibfnamefont {B.~R.}\ \bibnamefont {Ortiz}},
  \bibinfo {author} {\bibfnamefont {A.~K.}\ \bibnamefont {Hay}}, \bibinfo
  {author} {\bibfnamefont {S.~D.}\ \bibnamefont {Wilson}}, \bibinfo {author}
  {\bibfnamefont {R.}~\bibnamefont {Seshadri}},\ and\ \bibinfo {author}
  {\bibfnamefont {D.}~\bibnamefont {Johrendt}},\ }\bibfield  {title} {\bibinfo
  {title} {{The kagom{\'e} metals RbTi$_3$Bi$_5$ and CsTi$_3$Bi$_5$}},\
  }\href@noop {} {\bibfield  {journal} {\bibinfo  {journal} {Z. Naturforsch.
  B}\ }\textbf {\bibinfo {volume} {77}},\ \bibinfo {pages} {757} (\bibinfo
  {year} {2022})}\BibitemShut {NoStop}%
\bibitem [{\citenamefont {Yang}\ \emph {et~al.}(2022)\citenamefont {Yang},
  \citenamefont {Zhao}, \citenamefont {Yi}, \citenamefont {Liu}, \citenamefont
  {You}, \citenamefont {Zhang}, \citenamefont {Guo}, \citenamefont {Lin},
  \citenamefont {Shen}, \citenamefont {Chen} \emph
  {et~al.}}]{yang2022titanium}%
  \BibitemOpen
  \bibfield  {author} {\bibinfo {author} {\bibfnamefont {H.}~\bibnamefont
  {Yang}}, \bibinfo {author} {\bibfnamefont {Z.}~\bibnamefont {Zhao}}, \bibinfo
  {author} {\bibfnamefont {X.-W.}\ \bibnamefont {Yi}}, \bibinfo {author}
  {\bibfnamefont {J.}~\bibnamefont {Liu}}, \bibinfo {author} {\bibfnamefont
  {J.-Y.}\ \bibnamefont {You}}, \bibinfo {author} {\bibfnamefont
  {Y.}~\bibnamefont {Zhang}}, \bibinfo {author} {\bibfnamefont
  {H.}~\bibnamefont {Guo}}, \bibinfo {author} {\bibfnamefont {X.}~\bibnamefont
  {Lin}}, \bibinfo {author} {\bibfnamefont {C.}~\bibnamefont {Shen}}, \bibinfo
  {author} {\bibfnamefont {H.}~\bibnamefont {Chen}}, \emph {et~al.},\
  }\bibfield  {title} {\bibinfo {title} {{Titanium-based kagome superconductor
  CsTi$_3$Bi$_5$ and topological states}},\ }\href@noop {} {\bibfield
  {journal} {\bibinfo  {journal} {arXiv preprint arXiv:2209.03840}\ } (\bibinfo
  {year} {2022})}\BibitemShut {NoStop}%
\bibitem [{\citenamefont {Bie}\ \emph {et~al.}(2007)\citenamefont {Bie},
  \citenamefont {Moore}, \citenamefont {Piercey}, \citenamefont {Tkachuk},
  \citenamefont {Zelinska},\ and\ \citenamefont {Mar}}]{bie2007ternary}%
  \BibitemOpen
  \bibfield  {author} {\bibinfo {author} {\bibfnamefont {H.}~\bibnamefont
  {Bie}}, \bibinfo {author} {\bibfnamefont {S.~D.}\ \bibnamefont {Moore}},
  \bibinfo {author} {\bibfnamefont {D.~G.}\ \bibnamefont {Piercey}}, \bibinfo
  {author} {\bibfnamefont {A.~V.}\ \bibnamefont {Tkachuk}}, \bibinfo {author}
  {\bibfnamefont {O.~Y.}\ \bibnamefont {Zelinska}},\ and\ \bibinfo {author}
  {\bibfnamefont {A.}~\bibnamefont {Mar}},\ }\bibfield  {title} {\bibinfo
  {title} {{Ternary rare-earth titanium antimonides: phase equilibria in the
  RE--Ti--Sb (RE= La, Er) systems and crystal structures of
  RE$_2$Ti$_7$Sb$_{12}$ (RE= La, Ce, Pr, Nd) and RETi$_3$(Sn$_x$Sb$_{1-x}$)$_4$
  (RE= Nd, Sm)}},\ }\href@noop {} {\bibfield  {journal} {\bibinfo  {journal}
  {J. Solid State Chem."}\ }\textbf {\bibinfo {volume} {180}},\ \bibinfo
  {pages} {2216} (\bibinfo {year} {2007})}\BibitemShut {NoStop}%
\bibitem [{ESI()}]{ESI}%
  \BibitemOpen
  \href@noop {} {}\bibinfo {note} {See Supplemental Information for further
  details}\BibitemShut {NoStop}%
\bibitem [{\citenamefont {Oszl{\'a}nyi}\ and\ \citenamefont
  {S{\"u}t{\H{o}}}(2004)}]{oszlanyi2004ab}%
  \BibitemOpen
  \bibfield  {author} {\bibinfo {author} {\bibfnamefont {G.}~\bibnamefont
  {Oszl{\'a}nyi}}\ and\ \bibinfo {author} {\bibfnamefont {A.}~\bibnamefont
  {S{\"u}t{\H{o}}}},\ }\bibfield  {title} {\bibinfo {title} {Ab initio
  structure solution by charge flipping},\ }\href@noop {} {\bibfield  {journal}
  {\bibinfo  {journal} {Acta Crystallogr. A}\ }\textbf {\bibinfo {volume}
  {60}},\ \bibinfo {pages} {134} (\bibinfo {year} {2004})}\BibitemShut
  {NoStop}%
\bibitem [{\citenamefont {Oszl{\'a}nyi}\ and\ \citenamefont
  {S{\"u}t{\H{o}}}(2005)}]{oszlanyi2005ab}%
  \BibitemOpen
  \bibfield  {author} {\bibinfo {author} {\bibfnamefont {G.}~\bibnamefont
  {Oszl{\'a}nyi}}\ and\ \bibinfo {author} {\bibfnamefont {A.}~\bibnamefont
  {S{\"u}t{\H{o}}}},\ }\bibfield  {title} {\bibinfo {title} {{Ab initio
  structure solution by charge flipping. II. Use of weak reflections}},\
  }\href@noop {} {\bibfield  {journal} {\bibinfo  {journal} {Acta Crystallogr.
  A}\ }\textbf {\bibinfo {volume} {61}},\ \bibinfo {pages} {147} (\bibinfo
  {year} {2005})}\BibitemShut {NoStop}%
\bibitem [{\citenamefont {Coelho}(2007)}]{coelho2007charge}%
  \BibitemOpen
  \bibfield  {author} {\bibinfo {author} {\bibfnamefont {A.}~\bibnamefont
  {Coelho}},\ }\bibfield  {title} {\bibinfo {title} {A charge-flipping
  algorithm incorporating the tangent formula for solving difficult
  structures},\ }\href@noop {} {\bibfield  {journal} {\bibinfo  {journal} {Acta
  Crystallogr. A}\ }\textbf {\bibinfo {volume} {63}},\ \bibinfo {pages} {400}
  (\bibinfo {year} {2007})}\BibitemShut {NoStop}%
\bibitem [{\citenamefont {Coelho}(2018)}]{Coelho}%
  \BibitemOpen
  \bibfield  {author} {\bibinfo {author} {\bibfnamefont {A.~A.}\ \bibnamefont
  {Coelho}},\ }\bibfield  {title} {\bibinfo {title} {{{\it TOPAS} and {\it
  TOPAS-Academic}: an optimization program integrating computer algebra and
  crystallographic objects written in C++}},\ }\href
  {https://doi.org/10.1107/S1600576718000183} {\bibfield  {journal} {\bibinfo
  {journal} {J. Appl. Crystallogr.}\ }\textbf {\bibinfo {volume} {51}},\
  \bibinfo {pages} {210} (\bibinfo {year} {2018})}\BibitemShut {NoStop}%
\bibitem [{\citenamefont {Sheldrick}(2008)}]{sheldrick2008short}%
  \BibitemOpen
  \bibfield  {author} {\bibinfo {author} {\bibfnamefont {G.~M.}\ \bibnamefont
  {Sheldrick}},\ }\bibfield  {title} {\bibinfo {title} {{A short history of
  SHELX}},\ }\href@noop {} {\bibfield  {journal} {\bibinfo  {journal} {Acta
  Crystallogr. A}\ }\textbf {\bibinfo {volume} {64}},\ \bibinfo {pages} {112}
  (\bibinfo {year} {2008})}\BibitemShut {NoStop}%
\bibitem [{\citenamefont {Kresse}\ and\ \citenamefont
  {Furthm{\"u}ller}(1996{\natexlab{a}})}]{kresse1996efficient}%
  \BibitemOpen
  \bibfield  {author} {\bibinfo {author} {\bibfnamefont {G.}~\bibnamefont
  {Kresse}}\ and\ \bibinfo {author} {\bibfnamefont {J.}~\bibnamefont
  {Furthm{\"u}ller}},\ }\bibfield  {title} {\bibinfo {title} {Efficient
  iterative schemes for ab initio total-energy calculations using a plane-wave
  basis set},\ }\href@noop {} {\bibfield  {journal} {\bibinfo  {journal} {Phys.
  Rev. B}\ }\textbf {\bibinfo {volume} {54}},\ \bibinfo {pages} {11169}
  (\bibinfo {year} {1996}{\natexlab{a}})}\BibitemShut {NoStop}%
\bibitem [{\citenamefont {Kresse}\ and\ \citenamefont
  {Furthm{\"u}ller}(1996{\natexlab{b}})}]{kresse1996efficiency}%
  \BibitemOpen
  \bibfield  {author} {\bibinfo {author} {\bibfnamefont {G.}~\bibnamefont
  {Kresse}}\ and\ \bibinfo {author} {\bibfnamefont {J.}~\bibnamefont
  {Furthm{\"u}ller}},\ }\bibfield  {title} {\bibinfo {title} {Efficiency of
  ab-initio total energy calculations for metals and semiconductors using a
  plane-wave basis set},\ }\href@noop {} {\bibfield  {journal} {\bibinfo
  {journal} {Comput. Mater. Sci.}\ }\textbf {\bibinfo {volume} {6}},\ \bibinfo
  {pages} {15} (\bibinfo {year} {1996}{\natexlab{b}})}\BibitemShut {NoStop}%
\bibitem [{\citenamefont {Bl{\"o}chl}(1994)}]{blochl1994projector}%
  \BibitemOpen
  \bibfield  {author} {\bibinfo {author} {\bibfnamefont {P.~E.}\ \bibnamefont
  {Bl{\"o}chl}},\ }\bibfield  {title} {\bibinfo {title} {Projector
  augmented-wave method},\ }\href@noop {} {\bibfield  {journal} {\bibinfo
  {journal} {Phys. Rev. B}\ }\textbf {\bibinfo {volume} {50}},\ \bibinfo
  {pages} {17953} (\bibinfo {year} {1994})}\BibitemShut {NoStop}%
\bibitem [{\citenamefont {Kresse}\ and\ \citenamefont
  {Joubert}(1999)}]{kresse1999ultrasoft}%
  \BibitemOpen
  \bibfield  {author} {\bibinfo {author} {\bibfnamefont {G.}~\bibnamefont
  {Kresse}}\ and\ \bibinfo {author} {\bibfnamefont {D.}~\bibnamefont
  {Joubert}},\ }\bibfield  {title} {\bibinfo {title} {From ultrasoft
  pseudopotentials to the projector augmented-wave method},\ }\href@noop {}
  {\bibfield  {journal} {\bibinfo  {journal} {Phys. Rev. B}\ }\textbf {\bibinfo
  {volume} {59}},\ \bibinfo {pages} {1758} (\bibinfo {year}
  {1999})}\BibitemShut {NoStop}%
\bibitem [{\citenamefont {Setyawan}\ and\ \citenamefont
  {Curtarolo}(2010)}]{setyawan2010high}%
  \BibitemOpen
  \bibfield  {author} {\bibinfo {author} {\bibfnamefont {W.}~\bibnamefont
  {Setyawan}}\ and\ \bibinfo {author} {\bibfnamefont {S.}~\bibnamefont
  {Curtarolo}},\ }\bibfield  {title} {\bibinfo {title} {High-throughput
  electronic band structure calculations: Challenges and tools},\ }\href@noop
  {} {\bibfield  {journal} {\bibinfo  {journal} {Comput. Mater. Sci.}\ }\textbf
  {\bibinfo {volume} {49}},\ \bibinfo {pages} {299} (\bibinfo {year}
  {2010})}\BibitemShut {NoStop}%
\bibitem [{\citenamefont {Motoyama}\ \emph {et~al.}(2018)\citenamefont
  {Motoyama}, \citenamefont {Sezaki}, \citenamefont {Gouchi}, \citenamefont
  {Miyoshi}, \citenamefont {Nishigori}, \citenamefont {Mutou}, \citenamefont
  {Fujiwara},\ and\ \citenamefont {Uwatoko}}]{motoyama2018magnetic}%
  \BibitemOpen
  \bibfield  {author} {\bibinfo {author} {\bibfnamefont {G.}~\bibnamefont
  {Motoyama}}, \bibinfo {author} {\bibfnamefont {M.}~\bibnamefont {Sezaki}},
  \bibinfo {author} {\bibfnamefont {J.}~\bibnamefont {Gouchi}}, \bibinfo
  {author} {\bibfnamefont {K.}~\bibnamefont {Miyoshi}}, \bibinfo {author}
  {\bibfnamefont {S.}~\bibnamefont {Nishigori}}, \bibinfo {author}
  {\bibfnamefont {T.}~\bibnamefont {Mutou}}, \bibinfo {author} {\bibfnamefont
  {K.}~\bibnamefont {Fujiwara}},\ and\ \bibinfo {author} {\bibfnamefont
  {Y.}~\bibnamefont {Uwatoko}},\ }\bibfield  {title} {\bibinfo {title}
  {{Magnetic properties of new antiferromagnetic heavy-fermion compounds,
  Ce$_3$TiBi$_5$ and CeTi$_3$Bi$_4$}},\ }\href@noop {} {\bibfield  {journal}
  {\bibinfo  {journal} {Physica B Condens.}\ }\textbf {\bibinfo {volume}
  {536}},\ \bibinfo {pages} {142} (\bibinfo {year} {2018})}\BibitemShut
  {NoStop}%
\end{thebibliography}%

\end{document}